\documentclass[useAMS,usenatbib]{mn2e}

\usepackage{epsfig}
\usepackage{color}
\usepackage{placeins}

\newcommand{\msun}{\mbox{$M_\odot$}} 
\newcommand{\Mpc}{\mbox{$\rm Mpc$}} 

\begin{document}
\title[Star Formation Histories in Barred Spiral Galaxies]
{Star Formation History in Barred Spiral Galaxies. AGN Feedback}

\author[Fid\`ele Robichaud et al.]
{Fid\`ele Robichaud,$^{1,2}$
David Williamson,$^{1,2}$
Hugo Martel,$^{1,2}$
Daisuke Kawata,$^3$\newauthor
and Sara L. Ellison$^4$\\
$^1$D\'epartement de physique, de g\'enie physique et d'optique,
Universit\'e Laval, Qu\'ebec, QC, G1V 0A6, Canada\\
$^2$Centre de Recherche en Astrophysique du Qu\'ebec, QC, Canada\\
$^3$Mullard Space Science Laboratory, University College London,
Holmbury St Mary, Dorking, Surrey, UK\\
$^4$Department of Physics and Astronomy, University of
Victoria, Victoria, BC, Canada}

\date{Accepted XXX. Received XXX; in original form XXX}

\pagerange{\pageref{firstpage}--\pageref{lastpage}} \pubyear{XXX}

\maketitle

\label{firstpage}

\begin{abstract}
We present a numerical study
of the impact of AGN accretion and feedback on
the star formation history of barred disc galaxies.
Our goal is to determine whether the effect of feedback is positive 
(enhanced star formation) or negative (quenched star formation), and
to what extent. We performed a series of 12 hydrodynamical simulations
of disc galaxies, 10 barred and 2 unbarred, with various initial
gas fractions and AGN feedback prescriptions.
In barred galaxies, gas is driven toward the
centre of the galaxy and causes a starburst, followed by a slow decay,
while in unbarred galaxies the SFR increases slowly and steadily.
AGN feedback suppresses star formation near the central black hole.
Gas is pushed away from the black hole, and collides head-on with inflowing
gas, forming a dense ring at a finite radius where star formation is
enhanced.
We conclude that both negative and positive feedback are present,
and these effects mostly cancel out. There is no
net quenching or enhancement in star formation, but rather a displacement
of the star formation sites
to larger radii. In unbarred galaxies, where
the density of the central gas is lower, quenching of star formation near
the black hole is more efficient, and enhancement of star formation
at larger radii is less efficient. As a result, negative feedback dominates.
Lowering the gas fraction reduces the star formation rate at all radii,
whether or not there is a bar or an AGN.
\end{abstract}


\section{INTRODUCTION}

The existence of tight relations between the mass of supermassive 
black holes (SBH) producing Active Galactic Nuclei (AGN), the velocity
dispersion of their hosts galaxies 
\citep{fm00,gebhardtetal00,gultekinetal09},
and the bulge stellar mass \citep{mh03,hr04,mcm13}
strongly suggests that AGN
and their host do not evolve independently.
AGN deposit large amounts of energy into the surrounding interstellar
medium (ISM), and this energy might be sufficiently large to
accelerate the gas above the escape velocity and generate a galactic wind
that will deposit energy and metal-enriched gas into the intergalactic
medium.
This greatly affects the star formation histories of the host galaxies.
Star formation in galaxies is a highly inefficient process.
The baryon density parameter $\Omega_{\rm b0}$, estimated from the 
cosmic abundance of light elements, is of order 0.04 \citep{kirkmanetal03}.
By contrast, estimates of the average stellar mass density give
$\rho_*\simeq5.6\times10^8\msun\,\Mpc^{-3}$, corresponding to a
density parameter $\Omega_*=0.004$
\citep{sp99,coleetal01,belletal03,perez08}. Hence, the universal
star formation efficiency is only 10\%.
Several processes could reduce the global star formation rates (SFRs)
in galaxies, such as gas heating
\citep{mo77,bl00,e00,bl06,dfo06,ml11,hqm12}, gas stripping
\citep{km08,bek09,benitez13}, and gas evaporation
\citep{tassisetal03,wl06,wc07,pm07,yoshidaetal07,duffyetal14,gk14},
but by limiting the amount of gas available
to form stars, galactic outflows are probably the most important feedback
mechanism in isolated galaxies \citep{vcbh05,dvs08,bower12,fg13,ps13}.

Galaxies of different masses are not equally inefficient in forming stars.
The stellar-mass/halo-mass ratio, $M_*/M_h$, peaks at a halo mass
$M_h=10^{12}\msun$, with a ratio $M_*/M_h\sim0.03$ \citep{bcw10,mosteretal10}.
At both lower and higher masses, the ratio decreases, suggesting that two
different processes might be involved.
Feedback by supernovae can power galactic winds in
low-mass galaxies
\citep{bensonetal03,moetal04,ps13,mayikaetal14,mcnaughtetal14}, 
but this process is inefficient at high masses.
As the total galaxy mass increases, the binding energy of 
the gas increases faster than the energy released collectively by supernovae
\citep{pmg07,gbm09}.
Therefore, an alternate source of energy, such as AGN feedback,
is needed to explain galactic winds in high-mass galaxies
\citep{bensonetal03,kg05,bower06,somervilleetal08,guoetal11,bower12}.

This basic {\it negative feedback scenario} can explain the observed
galactic outflows, the observed $M_*/M_h$ ratios in high-mass galaxies,
and possibly the observed relations between SBH mass, velocity dispersion,
and bulge stellar mass.
However, it relies on an efficient coupling between the
energy released by the AGN and the ISM gas. The energy deposited by the
AGN can either increase the gas temperature (thermal feedback), or
accelerate the gas to large velocities (kinetic feedback). Thermal feedback
is less efficient than kinetic feedback, because a fraction of the
thermal energy will be radiated away before the resulting pressure
gradients can accelerate the gas. The relative importance of
thermal vs. kinetic feedback is still debated (e.g. \citealt{baraietal14}).
Also, it has been suggested that in an inhomogeneous ISM,
AGN outflows will
follow the path of least resistance, forming elongated cavities
instead of spherical ones \citep{gbm09,crescietal15}.
In this case, a substantial amount of gas might remain inside the host galaxy,
and increasing the AGN luminosity will not increase the fraction
of gas being ejected, but rather the velocity of the
gas that is ejected.
If the gas is not entirely expelled from the galaxy,
the fraction of gas that remains in the galaxy could
then be affected {\it positively\/} by the AGN feedback.
The injection of energy and momentum into the
surrounding gas can generate shock waves that will trigger star formation
by compressing the gas into dense shells \citep{if12,crescietal15}.

Observations reveal the existence of two AGN accretions modes, a radiatively
efficient mode associated with low-mass central black holes and high
SFRs that would result from interactions and mergers
\citep{kauffmannetal03,bestetal05,jnb09,kh09,smolcic09,bh12,capeloetal15},
and a radiatively inefficient accretion mode associated with
high-mass central black holes and low SFRs that would result from secular
evolution
\citep{bestetal05,bestetal06,allenetal06,hec06,hec07,gro13}.
In both cases, the anti-correlation between black hole mass and
SFR is consistent with negative feedback.
\citet{ellisonetal15b} showed that the SFRs of AGN hosts depend critically
on the selection criteria, with radio-selected AGN having very low SFRs,
optically-selected AGN having marginally suppressed SFRs, and
mid-IR-selected AGN having elevated SFRs.

Several recent observational and numerical studies
suggest that both positive and negative feedback are present in AGN hosts,
though their relative importance is unclear.
\citet{balmaverdeetal16} considered a sample of 132 low-redshifts quasars
($z<1$) selected from the Sloan Digital Sky Survey, with corresponding
photometric data from {\it Herschel}. They divided their sample into
strong-outflow and weak-outflow galaxies, and found no evidence of the SFR
being lower in galaxies with stronger outflows, as the negative feedback
scenario would predict. To the contrary, the SFRs are comparable to or
even higher than the ones in weak-outflow galaxies.
\citet{crescietal15} studied a radio-quiet QSO at redshift $z=1.59$
using SINFONI NIR integral field spectroscopy. Their study reveals the
presence of a powerful, highly-anisotropic outflow that expels gas
from the host galaxy along a cavity surrounded by a compressed layer
of gas. Star formation is suppressed inside the cavity, while enhanced
in the surrounding layer.
\citet{carnianietal16} studied two QSOs at high redshift ($z\sim2.5$)
that exhibit very strong outflows, and found that the effect on star
formation was marginal, and affected only a small region within the
host galaxy. These authors point out that their results are consistent with
the positive + negative feedback model suggested by \citet{crescietal15}.
\citet{rjbg15} performed a numerical simulation of the evolution
of a massive galaxy hosting an AGN. Their simulation produces a significant
outflow, but the SFR is reduced by only a few percent, because star-forming
clouds are too dense to be affected significantly by AGN feedback.
These authors conclude that the effect of AGN feedback on the SFR
of high-redshift galaxies is marginal. In a very recent
zoom-in simulation of a barred galaxy in a cosmological context,
\citet{spinosoetal17} found that in the presence of AGN feedback, most of
the gas in the centre of the galaxy is promptly consumed by star
formation, while only a small fraction is accreted by the central black hole.

While AGN feedback can greatly affect the evolution of the host galaxy,
the structure and dynamical evolution of the galaxy will, in turn,
influence the growth of the AGN.
In order to produce a luminous AGN, a galaxy requires a ready source
of gas, which may be reduced by star formation and
galactic outflows, Also, the galaxy must have the ability to drive this
gas toward the centre. This latter effect is particularly important
in barred galaxies.
In the presence of a bar, gas loses angular momentum and falls
toward the centre of the galaxy, where
it tends to form an elongated orbit inside the stellar bar 
(\citealt{athanassoula92,ce93,mtss02,mke13},
hereafter Paper~I).
If the gas eventually reaches the centre of the galaxy, it might
accrete onto the central black hole and fuel the AGN 
\citep{sfb89,sn93,hs94,combes03,jogee06}. 
Several observational studies find a higher 
bar fraction in AGN-host galaxies than non-AGN ones 
\citep{arsenault89,knapenetal00,laineetal02,gallowayetal15}.
However, other studies do not find any significant difference 
\citep{mmp95,mr95,mr97,hfs97,lsb04,haoetal09,ba09,leeetal12,cheungetal14},
or find a difference mostly in blue galaxies \citep{ooy12}.
Whilst changes in metallicity and SFRs are widely supported by
observations, the link between bars and AGN fueling has remained contentious.
Several observations reveal enhanced star formation in the central
regions of barred galaxies
\citep{hfs97,mf97,hm99,
emsellemetal01,knapenetal02,jogeeetal05,ellisonetal11}.
Gas converted into stars can no longer be accreted onto the central
black hole, and the relative importance of these two competing processes
is an unresolved issue \citep{ooy12,acl14,cisternasetal15}.

Despite conflicting observational results, in simulations, bars
represent a very effective way to move gas to the nuclear regions and hence
trigger an AGN. Therefore, simulations of barred galaxies represent
an excellent laboratory for the study of AGN feedback in galaxies.
We have performed a series of 10
chemodynamical simulations of isolated barred galaxies,
and for comparison, two simulations of isolated unbarred galaxies,
with various prescriptions for AGN feedback
Our goal is to determine the effect of AGN accretion and feedback
on the star formation history of barred spiral galaxies. In particular,
we want to determine if the AGN will deplete the gas reservoir
and quench star formation, or if the feedback effects of the AGN will
enhance star formation, and determine what is the timescale of those
events, compared to a galaxy of the same mass but with no AGN in its centre.

The remainder of this paper is as follows: in Section~2, we describe
our numerical algorithm, including our treatment of AGN feedback.
In Section~3, we present our suite of simulations.
Results are presented in Section~4 and discussed in
Section~5. Summary and conclusions are presented in Section~6.

\section{THE NUMERICAL METHOD}

\subsection{The GCD+ Algorithm}

All the simulations in this paper were performed using the numerical
algorithm GCD+ \citep{kg03,rk12,bkw12,
kawataetal13,kawataetal14}. GCD+ is a three-dimensional
tree/smoothed particle hydrodynamics (SPH) algorithm 
which simulates galactic chemodynamical evolution, accounting for
hydrodynamics, self-gravitation, star formation, supernova feedback,
metal enrichment and diffusion, and radiative cooling. 
Star formation
is handled by transforming gas particles into star particles:
if the local velocity of the gas particles
is convergent and one of them exceeds a given density threshold
$n_{\rm th}^{\phantom1}$, the gas particle may transform into a star particle
with a probability weighted by its density. The star particles have their
mass distributed accordingly to the \citet{salpeter_luminosity_1955}
initial mass function, and the metal enrichment they produce from Type~II 
and Ia supernovae is tabulated from \citet{woosley_evolution_1995} 
and \citet{iwamoto_nucleosynthesis_1999}.

Four main parameters govern the star formation rate and the supernovae
feedback \citep{rk12} and are fixed as follows: the
supernova energy output $E_{\mathrm{SN}}=1\times10^{50}\mathrm{erg}$,
the stellar wind energy output 
$E_{\mathrm{SW}}=5.0\times 10^{36}\rm erg\,s^{-1}$,
the star formation efficiency $C_*=0.02$, and the star formation density
threshold $n_{\rm th}^{\phantom1}=1\,\rm cm^{-3}$.

Cooling rates under the influence of a cosmological UV background are
calculated with the CLOUDY spectral synthesis code
\citep{ferlandetal98} with the assumption that gas is optically thin,
and are tabulated by redshift, density, and temperature for use in the GCD+
code. For the simulations presented in this paper,
we use the tables corresponding to redshift $z=0$.
Cooling is permitted down to $30\rm K$.
The gravitational softening length is fixed at $90\,\rm pc$. 
Gas particles have individual smoothing lengths, which are calculated
through an iterative procedure so that each gas particle has $\approx58$
neighbours. However, we impose a minimum smoothing length of $90\,\rm pc$.
The smoothing length of the particle representing the central black hole
is adjusted such that the black hole has $\approx70$ particles within its
zone of influence.

\subsection{AGN Feedback \& Dynamics}

Comparative studies of AGN feedback algorithms in major merger simulations 
\citep{wt13a,thackeretal14} demonstrate that AGN feedback
algorithms greatly differ in their accretion rates, and in the strength and
effects of their feedback. These models are generally based on the Bondi
accretion rate \citep{hl39,bh44,bondi52} for accretion onto a dense object,
\begin{equation}
\dot{M}_\mathrm{Bondi}=\frac{2\pi G^2 M_\mathrm{BH}^2 \rho_\infty}
{(c_\infty^2+v^2)^{3/2}},
\label{bondi}
\end{equation}

\noindent
where $M_\mathrm{BH}$ is the mass of the black hole, $\rho_\infty$ and $c_\infty$
are the density and sound-speed of the gas at infinity,
respectively, and $v$ is the speed
of the black hole relative to this distant gas.
Our algorithm is based on the `WT' model of \citet{wt13a}

The mass of the black hole is represented by two values: the dynamical mass
of the black hole particle ($M_\mathrm{dyn}$), and the internal
``sub-grid-scale'' mass of the physical black hole ($M_\mathrm{SGS}$).
The sub-grid-scale mass is used to calculate the accretion rate
in equation~(\ref{bondi}) and equation~(\ref{eddington}) below, while the
dynamical mass is used in kinematic interactions with other particles, and
in determining the particle's motion.
\textsc{GCD+} requires that gas and star particles have a constant mass,
and so to maintain mass conservation, the dynamical mass increases
discretely when an entire particle is accreted, while the sub-grid-scale
mass increases continually, as detailed below.

In numerical simulations, the mass of star particles can be a large enough
fraction of the black hole particle's mass that two-body scattering can
become significant. To reduce this source of error, we apply a damping
force to the black hole particle with a somewhat arbitrary time-scale of
$1\,\rm Myr$. Specifically, every time-step we apply the transform
${\bf v}_{\rm BH}^{\phantom i}\rightarrow
{\bf v}_{\rm BH}\exp[-\Delta t_{\rm BH}^{\phantom i}/({1\,\rm Myr})]$,
where ${\bf v}_{\rm BH}^{\phantom i}$ is the black hole particle's
velocity vector, and $\Delta t_{\rm BH}^{\phantom i}$ is its time-step.
This allows the black hole to sink
to the minimum of the potential well without being sensitive to two-body
kicks. We also force the black hole particle to have a time-step equal to
the smallest time-step in the galaxy, or $5\times10^4\,\rm yr$,
whichever is smallest. Finally, in some sets of initial conditions
(Runs~G and~I below), we place the black hole at a point
$\rm(1\,pc, 1\,pc, 1\,pc)$ from the centre of mass of the galaxy, so that
it is not too close to another particle. The black hole particle then
quickly finds the minimum of the potential well.

The Bondi accretion rate given by
equation~(\ref{bondi}) uses values at infinity.
As is typical in SPH simulations, we replace $\rho_\infty$ and
$c_\infty$ by the values calculated using the particles inside the
black hole's smoothing length, and we replace
$v$ by the speed of the black hole particle relative to the
SPH-smoothed gas velocity at its location. Finally, we use
for $M_\mathrm{BH}$ the internal mass of the black hole $M_\mathrm{SGS}$.
The smoothing length of the black hole is variable, as with the gas
particles, calibrated so that the black hole particle would have
$\sim92$~neighbouring gas particles if the gas was uniformly distributed.
In practice, the flat geometry of the gas near the black hole particle
causes the black hole particle to have $\sim70$~neighbour particles,
still a little above the median number of neighbours for a gas
particle. This larger number of neighbours reduces rapid variations in the
black hole's accretion rate.

The maximum accretion rate in spherically symmetric hydrodynamic equilibrium
is the Eddington accretion rate,
\begin{equation}
\dot{M}_\mathrm{Edd}=\frac{4\pi G M_\mathrm{BH} m_p}
{\epsilon_r\sigma_T^{\phantom1}c},
\label{eddington}
\end{equation}

\noindent
where $m_p$ is the mass of a proton, $\epsilon_r$ is the radiative efficiency,
and $\sigma_T^{\phantom1}$ 
is the Thomson cross-section. The radiative efficiency is a
free parameter, which we set to $\epsilon_r=0.1$, following \citet{ss73}.
We also increase our numerical accretion rate by a factor $\alpha$, to
reflect the underestimation of sound-speeds that results from resolution
limits \citep{bs09}. Hence our numerical accretion rate is
\begin{equation}
\dot{M}_\mathrm{num} = 
\min(\dot{M}_\mathrm{Edd},\alpha\dot{M}_\mathrm{Bondi}).
\label{LAGN}
\end{equation}

\noindent
We select $\alpha=100$, following \citet{sdh05}.

During each timestep $\Delta t$, the accretion rate $\dot M_{\rm num}$
is calculated using equations~(\ref{bondi})--(\ref{LAGN}) with
$M_{\rm BH}=M_{\rm SCS}$. Then
the internal mass is augmented using
$M_{\rm SGS}\to M_{\rm SGS}+\Delta M$, where
$\Delta M=\dot{M}_{\rm num}\Delta t$, to account for 
mass being accreted onto the black hole. However, the dynamical mass
$M_{\rm dyn}$, which enters into the calculation of gravity, is left unchanged,
and the mass of neighbouring particles is also left unchanged.
To ensure approximate agreement
between the dynamical mass and the internal mass of the black hole particle,
a gas particle is accreted onto the black hole particle when the internal
mass exceeds the dynamical mass by half of the mass of a gas particle.
This closest gas particle to the black hole particle is deleted,
and its mass is added to the dynamical mass of
the black hole particle. This insures
that $M_{\rm dyn}$ always remains within one half of a
gas particle mass of $M_{\rm SCS}$ during the course of
the simulation.\footnote{We note that the particles
being accreted are typically located at a distance from the black hole
comparable to or only slightly larger than the gravitational softening
length.}

A fraction $\epsilon_r=0.1$ of the accreted rest-mass energy
$\Delta Mc^2$ is returned in the form
of feedback. As the ISM is largely optically thin, only a portion of this
energy ($\epsilon_c=0.05$) is coupled to the ISM, giving a feedback energy of
\begin{equation}
\Delta E = \epsilon_r \epsilon_c \Delta M c^2.
\end{equation}

\noindent
This energy is divided evenly amongst all particles within the black hole
particle's smoothing length. A fraction, $f_\mathrm{th}$ of the energy is
applied as thermal energy, with the remaining applied directly as kinetic
energy by applying a radially directed momentum kick of 
$p=(1-f_\mathrm{th})c\Delta E/N_f$
to all $N_f$ particles within the black hole particle's smoothing length.
\citet{baraietal14} compared thermal feedback with direct input of
kinetic energy, and found that kinetic feedback has a stronger effect,
producing a clear outflow.

\section{THE SIMULATIONS}

\subsection{Initial Conditions}

For generating the initial conditions of our simulations, we use the same
technique as in \citet{gkc15} and~\citet{cmke16} (hereafter Paper~II).
Since our goal is to assess the importance of positive and negative
feedback on star formation, we examine galaxies where AGN feedback is strong,
but not so strong as to remove the gas from the galaxy and completely shut
off star formation.
The shape of the $M_h/M_*$ relation shows that SNe feedback
dominates for masses $M_*<3\times10^{10}\msun$, whereas most of the
gas in blown out for much larger masses \citep{bcw10,mosteretal10}.
For this reason, we selected an initial stellar
mass $M_*=5.80\times10^{10}\msun$, which is near the bottom of the
``AGN regime.'' The corresponding
halo mass is $M_{200}=2.306\times10^{12}\msun$.

We set up the stellar disc using an exponential surface density profile:
\begin{equation}
\rho_* = \frac{M_*}{4 \pi z_* R^2_*} {\rm sech}^2
\left({z\over z_*}\right)e^{-R/R_*}\,,
\label{eq:Stellardisk}
\end{equation}

\noindent
where $R_*$ is the scale length, $z_*$ is the
scale height, and $R$ and $z$ are the radial and horizontal coordinates
respectively. The gaseous disc has the same radial exponential surface
density, but its height is determined by imposing an initial hydrostatic
equilibrium within the gaseous disc \citep{sdh05}.
We then set an initial radial
metallicity profile in both the stellar and gaseous populations, with the
iron abundance being given by 
\begin{equation}
\label{eq:FeHprofile}
\mathrm{[Fe/H]}= 0.2 -0.05R,
\end{equation}

\noindent
where $R$ is in kpc. $\alpha-$elements are initially only present in the
stellar component and their abundance is given by 
\begin{equation}
\label{eq:alphaFeprofile}
[{\alpha}/{\rm Fe}]= -0.16[{\rm Fe/H}](R).
\end{equation}

We add to this value a gaussian scatter of 
$0.02\,\rm dex$ to create a local dispersion of the abundances. The star
particles are assigned an initial age using an age-metallicity relation 
$[\rm{Fe/H}]=-0.04\times{\rm age(Gyr)}$.
We do not use particles to represent the dark matter.
Instead, we assume a static halo with an NFW profile \citep{nfw96},
an approximation which is appropriate for simulations of isolated 
galaxies. This dark matter halo is characterised by a mass $M_{200}$
and a concentration parameter $c$.
We use two different values of $c$: 8 and 20.
As shown in Papers~I and II, the lower value allows the formation of a bar
by instability, while the larger value suppresses it. 

The scale length of the stellar disc is calculated using the relation 
between $R_{50}$, the half-light radius, and $M_*$, the mass of the 
stellar disc, as found by \citet{shenetal03}: 
\begin{equation}
R_{50}(\mathrm{kpc})=\gamma M_*^{\alpha} 
\left(1+\frac{M_*}{M_0}\right)^{\beta - \alpha},
\end{equation}

\noindent
where $\gamma$ is a scaling factor, $M_0$ is the characteristic mass of the 
transition between the low-mass behaviour as $M_*^{\alpha}$ and the high-mass 
one as $M_*^{\beta}$.
Continuing to follow \citet{shenetal03},
we use $\gamma=0.1$, $M_0=3.98\times10^{10}\msun$, 
$\alpha = 0.14$, and $\beta = 0.39$ to evaluate the half-light radius for a
given mass. Assuming that the half-light radius corresponds roughly with the 
half-mass radius, we integrate the density profile of the stellar disc 
(eq. \ref{eq:Stellardisk}) up 
to $R_{50}$ to obtain a transcendental relation between $R_*$ and $R_{50}$ 
which lets us compute the scale length from the disc mass.
For the mass considered in this paper,
this gives a scale length $R_*=2.4\,\rm kpc$.
We also set the scale height of the stellar disc to $z_*=0.48\,\rm kpc$.

We set the fiducial parameters for the gaseous disc using the 
same relation as \citet{coxetal06}, itself derived from \citet{belletal03}: 
\begin{equation}
\log M_{\rm gas}=0.78\log M_*-1.74,
\label{mgasinit}
\end{equation}

\noindent
where both masses are expressed in units of $10^{10}\msun$.
This gives a gas mass $M_{\rm gas}=1.38\times10^{10}\msun$, and an initial
gas fraction $f_{\rm gas}=M_{\rm gas}/(M_*+M_{\rm gas})=0.192$.
We will actually experiment with various values of $f_{\rm gas}$,
while keeping $M_*+M_{\rm gas}$ constant.
The scale length of the gas disc
is $R_{\rm gas}=4.79\,\rm kpc$.
To determine the vertical distribution of the gas, we follow the
prescription described in \citet{sdh05}. We first use CLOUDY to
determine an effective equation of state of the form $P=P(\rho)$.
The initial vertical distribution of the gas disc is then set so that it is in
hydrodynamic equilibrium. This produces a flared disc with a scale height
$z_{\rm gas}$ of $30\,\rm pc$ at the centre
and $600\,\rm pc$ at $R=40\,\rm kpc$.

Finally, we put the black hole initially at rest at the centre of mass of 
the galaxy, and initialise its masses at 
$M_{\rm dyn,i}=M_{\rm SGS,i}=10^6\msun$. The black hole particle 
is included even in simulations without AGN feedback, to prevent the small 
dynamical effects of the presence of a black hole particle from influencing 
the results.

\subsection{Runs and Parameters}

\begin{table}
\centering
\begin{minipage}{140mm}
\caption{Parameters of the simulations.}
\begin{tabular}{@{}lcccccr@{}}
\hline
Run & barred & $N_{\rm gas}{}^a$ & $N_{\rm star}{}^b$ & 
$f_{\rm gas}{}^c$ & $t_{\rm AGN}^{\phantom1}{}^d$ & $f_{\rm kin}{}^e$\\
\hline
 A & yes & 122694 & 514541 & 0.192 & $\cdots$ & $\cdots$ \\
 B & yes & 122694 & 514541 & 0.192 & 0        & 0 \\
 C & yes & 122694 & 514541 & 0.192 & 0        & 0.1 \\
 D & yes & 122694 & 514541 & 0.192 & 0        & 0.2 \\
 E & yes & 122694 & 514541 & 0.192 & 0.5      & 0.1 \\
 F & yes & 122694 & 514541 & 0.192 & 0.5      & 0.2 \\
 G & yes &  61347 & 575888 & 0.096 & $\cdots$ & $\cdots$ \\
 H & yes &  61347 & 575888 & 0.096 & 0        & 0.2 \\
 I & yes &  30674 & 606561 & 0.048 & $\cdots$ & $\cdots$ \\
 J & yes &  30674 & 606561 & 0.048 & 0        & 0.2 \\
 K & no  & 122694 & 514541 & 0.192 & $\cdots$ & $\cdots$ \\
 L & no  & 122694 & 514541 & 0.192 & 0        & 0.2 \\
\hline
\end{tabular}
\end{minipage}
\medskip
\begin{minipage}{84mm}
$^a$ Initial number of gas particles.\\
$^b$ Initial number of star particles.\\
$^c$ Initial gas fraction.
Galaxies with $f_{\rm gas}=0.192$, 0.096, and 0.048 are ``gas-normal,''
``gas-poor,'' and ``gas-very-poor,'' respectively.\\
$^d$ Time in Gyr when AGN accretion and feedback is turned on.\\
$^e$ Fraction of feedback energy applied as kinetic energy.
\end{minipage}
\end{table}

We performed a series of 12 simulations. Run~A is a simulation of
a barred galaxy without
an AGN. In Runs~B, C, and~D, we include an AGN, and vary the relative
strength of thermal and kinetic feedback.

In all simulations, we start with an axisymmetric disc, and the bar
forms by instability. However, there are other processes that can lead
to the formation of a bar.
To investigate the possibility that the
AGN turns on in a galaxy where the bar is already formed,
we have performed two simulations, E and F, where we delay the
turn-on of AGN feedback until the bar has formed,
at $t_{\rm AGN}^{\phantom1}=0.5\,\rm Gyr$.

Runs A--F all have an initial gas fraction $f_{\rm gas}=0.192$,
consistent with equation~(\ref{mgasinit}).
Runs~G and~I, are similar to Run~A (no AGN), and Runs~H and~J are
similar to Run~D (same AGN feedback prescription), 
but these runs start with a lower initial gas fraction.
Runs~K and L are similar to Runs~A and D, respectively, but we used
a concentration parameter $c=20$ to prevent the formation of a bar.
The parameters of the simulations are listed in Table~1.
$N_{\rm gas}$ and $N_{\rm star}$ are the initial number of gas and star 
particles respectively. $f_{\rm gas}$ is the initial gas fraction, 
$t_{\rm AGN}^{\phantom1}$ is the time when AGN
accretion and feedback is turned on, and $f_{\rm kin}=1-f_{\rm th}$ 
is the fraction of feedback energy applied as kinetic energy.
With these parameters, each particle has a mass of $8.60\times10^4\msun$.
We will refer to galaxies K, L, and
A through~F as ``gas-normal,'' galaxies G and~H as
``gas-poor,'' and galaxies I and~J as ``gas-very-poor.''

\FloatBarrier

\section{RESULTS}

\subsection{Bar Formation}

\begin{figure*}
\begin{center}
\includegraphics[scale=0.8]{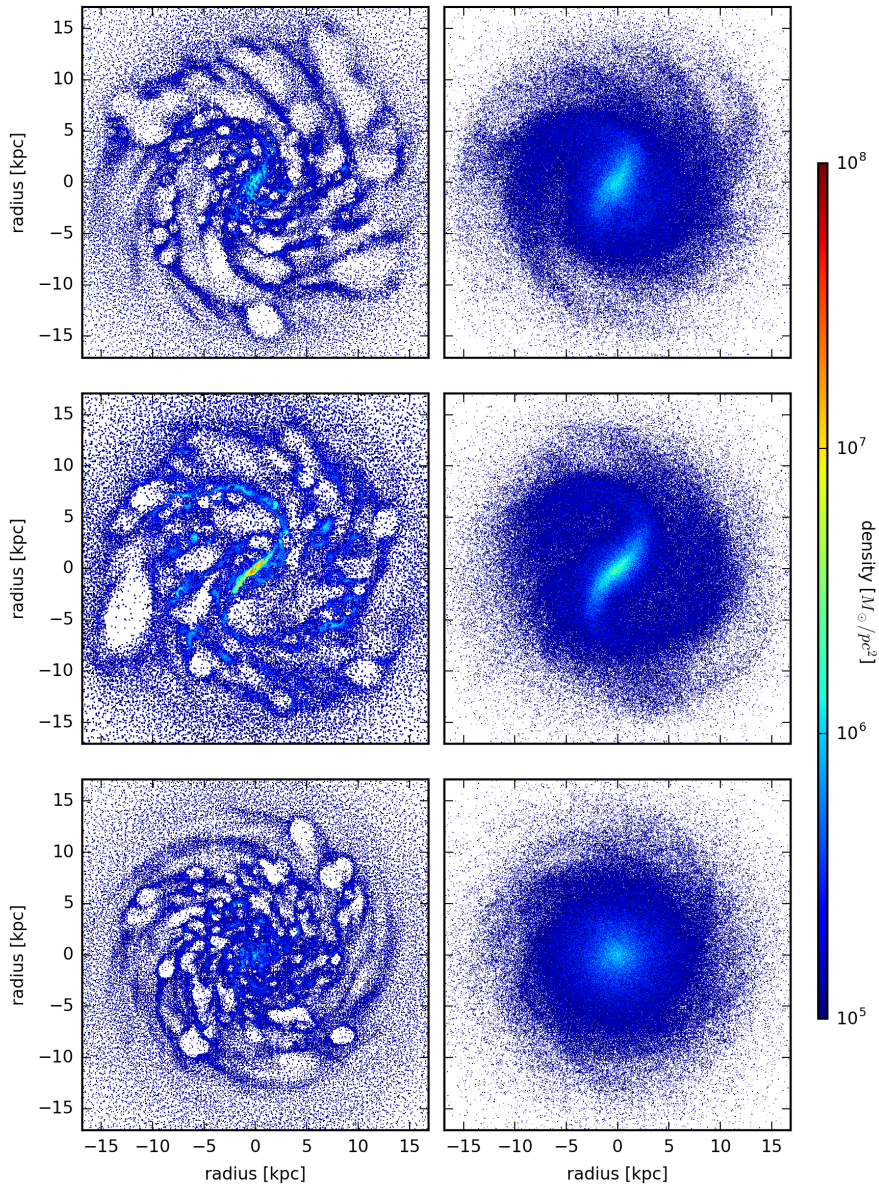}
\caption{Configuration of the system at $t=400\,\rm Myr$,
showing the column densities of gas (left) and stars (right).
Top row: Run~A (bar, no AGN); middle row: Run D (bar + AGN);
bottom row: Run~K (no AGN, no bar).}
\label{full}
\end{center}
\end{figure*}

\begin{figure*}
\begin{center}
\includegraphics[scale=0.8]{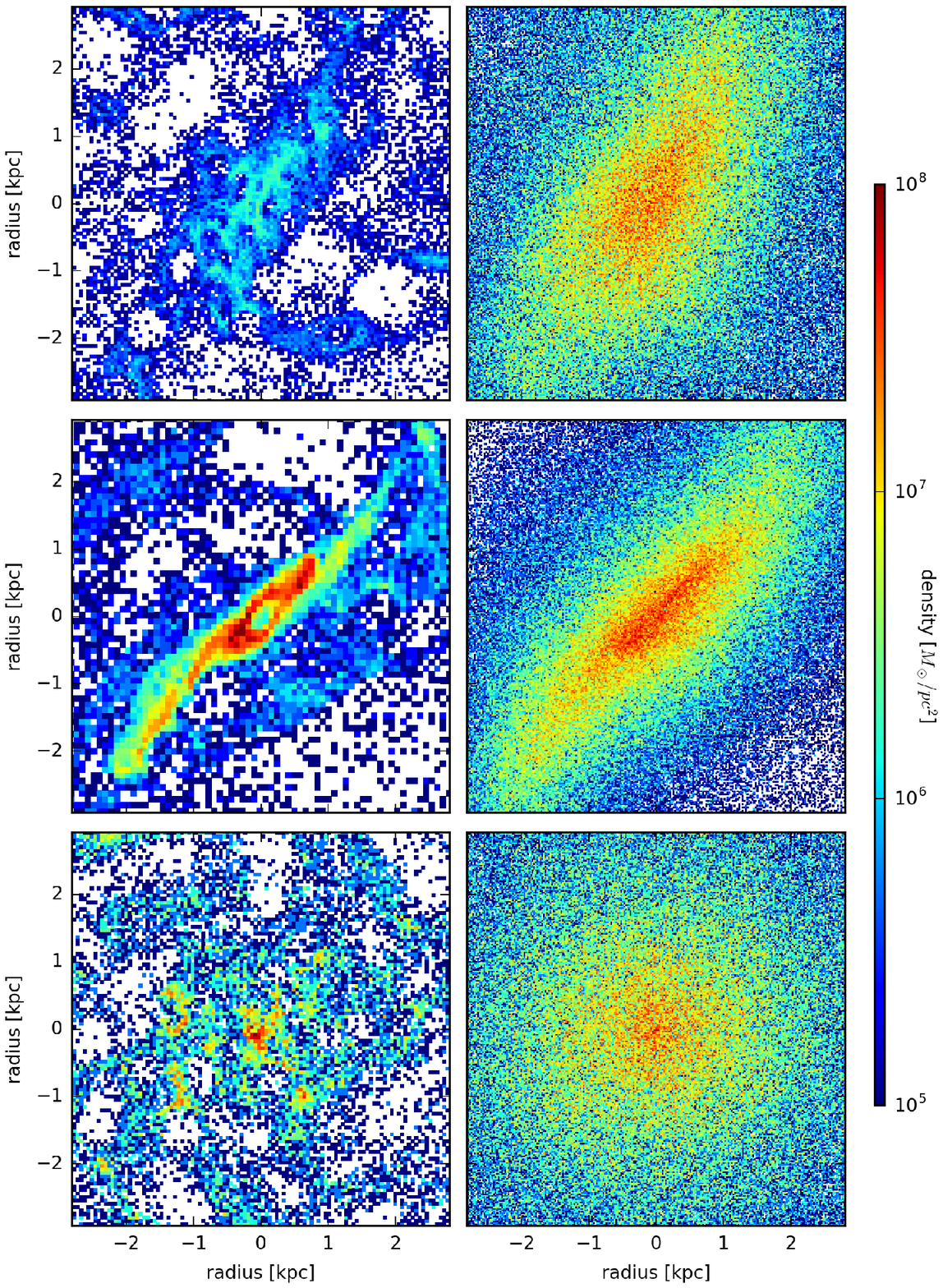}
\caption{Zoom-in of the regions shown in Fig.~\ref{full}.}
\label{zoom}
\end{center}
\end{figure*}

Figures~\ref{full} and \ref{zoom} show the column density of gas
(left panels) and stars (right panels) at $t=400\,\rm Myr$, for
Runs~A, D, and K. Figure~\ref{full} shows the entire galaxies, while
Figure~\ref{zoom} shows zoom-ins of the central regions.
Runs A and D differ by the presence of an AGN.
In both runs, we clearly see the bar and the pattern of spiral
arms, both in the gas and stars. However, at that particular
time the bar in Run~D is significantly
longer and more prominent than in Run~A,
because it forms earlier. Runs~A and~K differ by the presence
of a bar. The spiral pattern can be seen in the gas for Run~K, but it
is much less prominent than in the runs with a bar.
Not having a bar greatly reduces the gas flow
toward the centre of the galaxy, resulting in a much flatter surface
density gradient and a much lower central surface density, 
both in stars and gas. Results for Run~L are similar.
The numerous cavities
we see in the gas distributions are caused by SNe feedback.

\begin{figure}
\begin{center}
\includegraphics[scale=0.50]{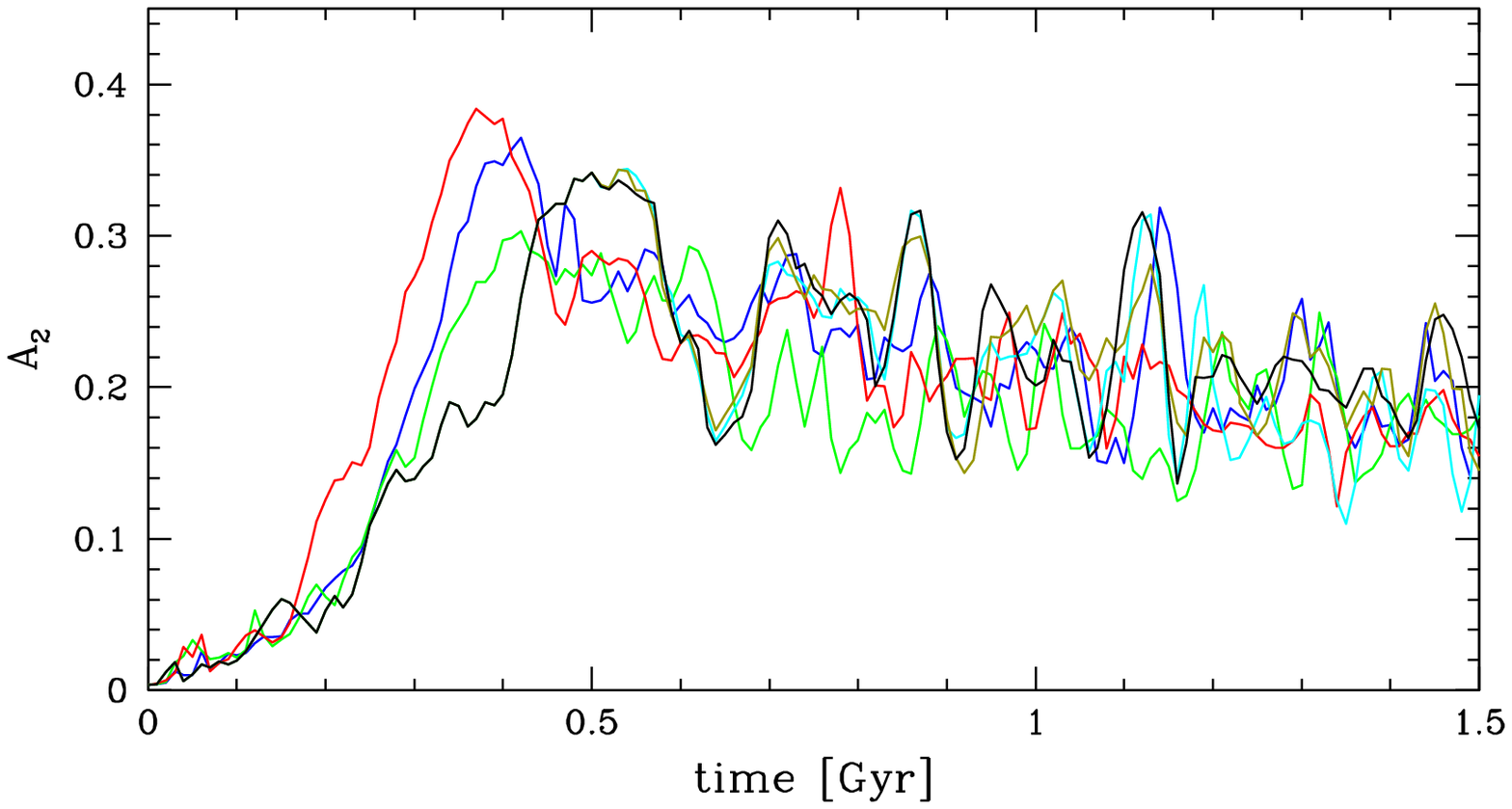}
\caption{
Bar strength vs. time for gas-normal galaxies. 
Black line: Run A (no AGN); blue line: Run B (thermal 
feedback); green line; Run C ($f_{\rm kin}=0.1$); red line:
Run D ($f_{\rm kin}=0.2$); cyan line: Run E (delayed AGN, $f_{\rm kin}=0.1$);
gold line: Run F (delayed AGN, $f_{\rm kin}=0.2$). Note that gold and cyan lines
are identical to black line at $t<0.5\,\rm Gyr$.}
\label{bar}
\end{center}
\end{figure}

As in Paper II, we calculate bar strength using a method proposed by
\citet{am02}, based on the components of the Fourier
decomposition of the azimuthal distribution of particles. 
In Figure \ref{bar} we present the evolution of the bar 
strength as a function of time for gas-normal galaxies. 
In the absence of AGN feedback (Run A, black line), the bar strength
increases steadily, reaching a peak value of $A_2=0.34$ by 
$t=500\,\rm Myr$. Afterward, the value of $A_2$ strongly oscillates,
with a net decrease over time, down to $A_2\sim0.2$ at
$t=1.5\,\rm Gyr$. With AGN feedback included (Runs B, C, and D),
the bar grows faster and the peak value of $A_2$ is reached earlier,
after which the evolution is similar to Run A.
In all simulations, the initial disc is in a state of unstable
equilibrium. AGN feedback seems to hasten the growth of the instability,
possibly by perturbing the equilibrium earlier. A similar effect was noticed
by \citet{spinosoetal17}.
Even though Run C is an
intermediate between Runs B and D, it has the lowest peak value:
$A_2=0.30$, compared to 0.36 for Run B and 0.38 for Run D.
Such non-monotonicity is not unexpected, considering again
that the bar in these simulations forms by instability.
In Runs~E and F, we turn on AGN accretion
and feedback at $t=500\,\rm Myr$,
which coincidently corresponds to the moment when the bar
has reached its maximum strength. Adding feedback this late does not
significantly affect the subsequent evolution of the bar.

\subsection{Central Gas Reservoir}

\begin{figure}
\begin{center}
\includegraphics[scale=0.50]{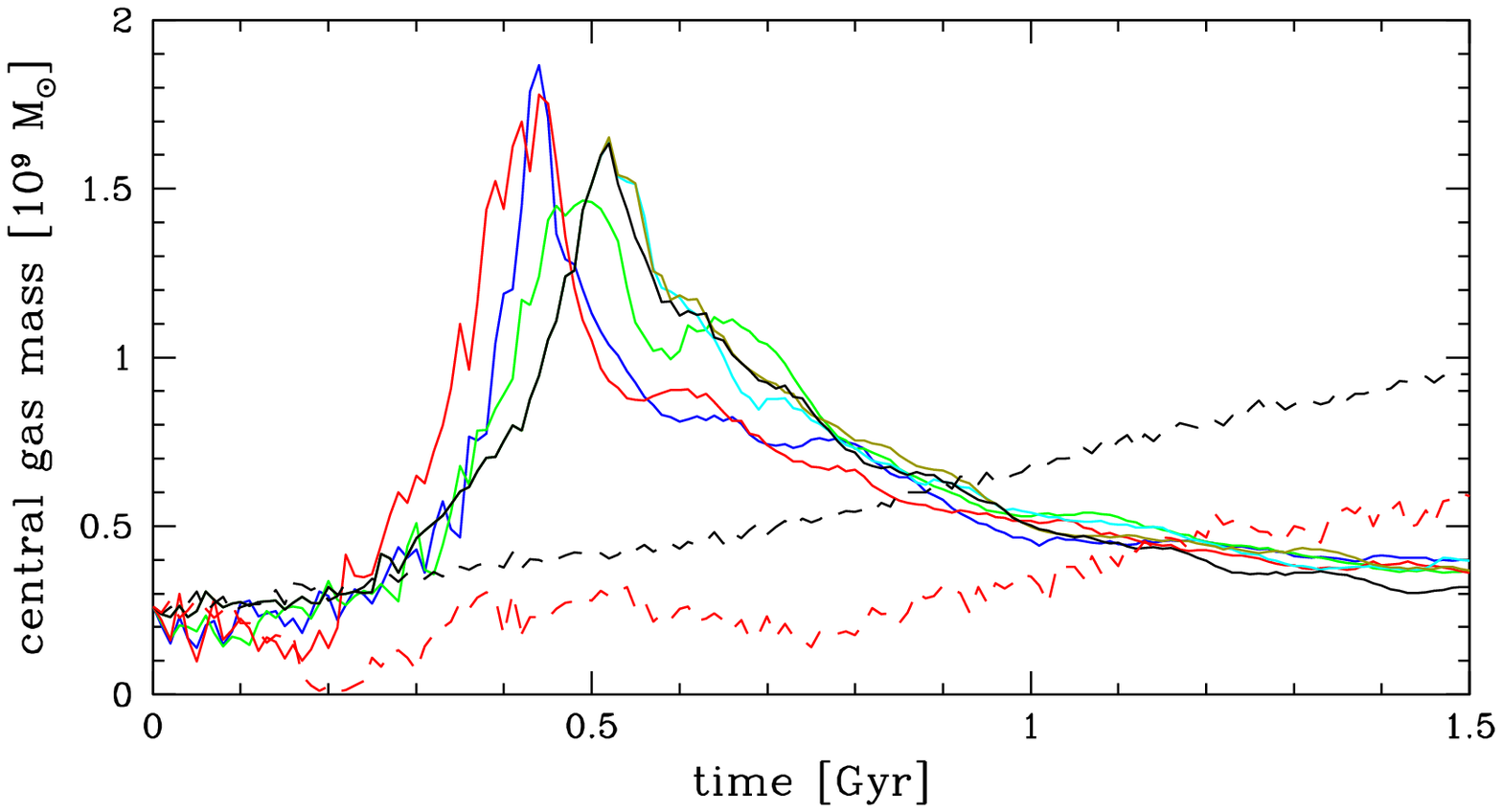}
\caption{
Gas mass in the central region vs. time for gas-normal galaxies.
Solid black lines: Run A (No AGN); blue lines: Run B (thermal 
feedback); green lines; Run C ($f_{\rm kin}=0.1$); solid red lines:
Run D ($f_{\rm kin}=0.2$); cyan lines: Run E (delayed AGN, $f_{\rm kin}=0.1$);
gold lines: Run F (delayed AGN, $f_{\rm kin}=0.2$);
dashed black lines: Run K (no bar, no AGN);
dashed red lines: Run L (no bar, $f_{\rm kin}=0.2$).
Note that gold and cyan lines
are identical to black line at $t<0.5\,\rm Gyr$.}
\label{gas_centre}
\end{center}
\end{figure}

Figure~\ref{gas_centre} shows the evolution of the
gas mass in the central $1\,\rm kpc$,
for gas-normal galaxies.\footnote{The {\it centre}
of the galaxy is defined as the centre of the
NFW dark matter halo. As we explained in \S~2.1, the black hole is allowed
to move, and can be slightly displaced relative to the centre.}
Comparing the location and height of the peaks
in the top panel with the ones in Figure~\ref{bar}, we see that
the early evolution of the central gas mass almost perfectly
mirrors the early evolution of the bar strength. The bar forms earlier
in simulations with AGN feedback, allowing the central gas density to build
up more rapidly.
Afterward, at
$t>500\,\rm Myr$, the mass of gas decreases while the bar
remains present. 
After $t=800\,\rm Myr$,
the evolution is essentially the same for all runs, indicating that
AGN accretion and feedback has little effect in the 
central $1\,\rm kpc$ once
most of the central gas has been either accreted, converted to stars
or pushed out to larger radius. In barred galaxies, the central gas mass
rapidly peaks because of bar-induced gas inflow, and then drops rapidly.
This reduction in central gas mass is caused primarily
by gas being converted to stars, as we will show below.
The situation
is drastically different in unbarred galaxies. Without feedback
(Run~K, dashed black line),
the gas mass in the central $1\,\rm kpc$ region increases
slowly and monotonically, with a slight change of slope at $t\sim900\,\rm Myr$.
With feedback (Run~L, dashed red line),
the growth is even slower and also more irregular, but does
not have any significant peak.

\subsection{Black Hole Growth and Feedback}

\begin{figure}
\begin{center}
\includegraphics[scale=0.50]{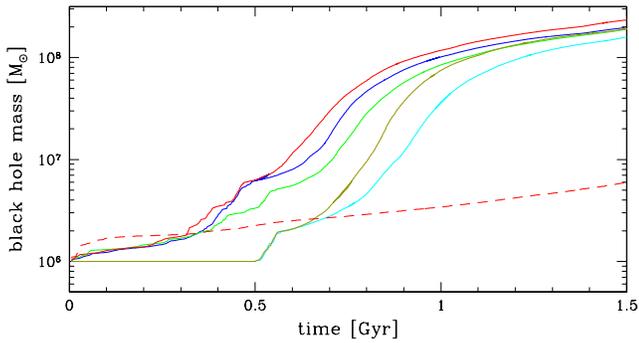}
\caption{
Black hole mass vs. time for gas-normal galaxies. Blue line: Run B (thermal 
feedback); green line; Run C ($f_{\rm kin}=0.1$); solid red line:
Run D ($f_{\rm kin}=0.2$); cyan line: Run E (delayed AGN, $f_{\rm kin}=0.1$);
gold line: Run F (delayed AGN, $f_{\rm kin}=0.2$);
dashed red line: Run L (no bar, $f_{\rm kin}=0.2$). Note that gold and cyan
lines are identical at $t<0.5\,\rm Gyr$.}
\label{mass_BH}
\end{center}
\end{figure}

Figure~\ref{mass_BH} shows the black hole mass $M_{\rm BH}$ vs. time,
for all gas-normal galaxies except Runs~A and K, which contain
no AGN.\footnote{Unless 
specified otherwise, the black hole
mass $M_{\rm BH}$ refers to the sub-grid mass $M_{\rm SGS}$, and
not the dynamical mass $M_{\rm dyn}$ (see Section 2.2).} 
For the first $300\,\rm Myr$, the black holes
grow at essentially the same rate in Runs B, C, and D. During that
period, AGN feedback is too weak to significantly affect the accretion
rate onto the black hole. $t=300\,\rm Myr$ is about the time when the
bar strength reaches $A_2=0.2$. As we showed in Section~4.2, a 
stronger bar drives larger amounts
of gas toward the centre, increasing both the accretion rate and the
strength of the feedback. The subsequent evolution differs for Runs
B, C, and D. The difference in black hole mass between
Runs C and D reaches a factor of
2.5 around $t=700\,\rm Myr$, but by $t=1.5\,\rm Gyr$ the differences
are less than 50\%. We find again a non-monotonicity
in the results: Run C, which has the intermediate kinetic feedback 
($f_{\rm kin}=0.1$), lies below Runs B and D and not between them.
The time when the bar forms and drives gas toward the centre appears
more critical than the nature of the feedback in determining
the growth rate of a black hole.

In Runs E and F, accretion and feedback are turned off until
$t=500\,\rm Myr$. Once it is turned on, the black hole mass 
first grows very
rapidly, because a large amount of gas has been accumulated in the
centre, waiting to be accreted. Initially, the mass growth is the same in
both runs, and the black holes double their masses in less
than $50\,\rm Myr$.
They start to differ once feedback becomes important, and from
there the evolution is qualitatively similar to Runs B, C, and D.
In all five runs, feedback starts to make a difference once the black hole
reaches a mass of order $M_{\rm BH}\sim2\times10^6\msun$, that is,
double its initial mass. For Runs B, C, and D, this coincides with the
time when the bar reaches its maximum strength.

The dashed red line in Figure~\ref{mass_BH} shows the growth of the black
hole mass for Run~L. In the absence of bar-driven gas inflow, the black hole
mass increases slowly and steadily. By $1\,\rm Gyr$ its mass has increased
by a factor of 6, compared to 234 for Run~D. 
Interestingly, there is a sudden increase in mass at the beginning of the 
simulation for Run~L, where the mass of the black 
hole goes up by 50\% after a mere $25\,\rm Myr$.
We also see a weaker initial increase in Runs B, C, and D.
In all cases, the black hole first accretes the gas located in its
vicinity, and then gas flowing from larger distances.
The higher concentration parameter in Run L
increases the gas density and the depth
of the potential at the centre, driving more cooling and more accretion.
Afterward, the accretion rate is reduced in Run~L
because less gas is being driven toward the centre, and by 
$t=300\,\rm Myr$, Runs~B, C, and D have caught up with it.

\begin{figure*}
\begin{center}
\includegraphics[scale=0.75]{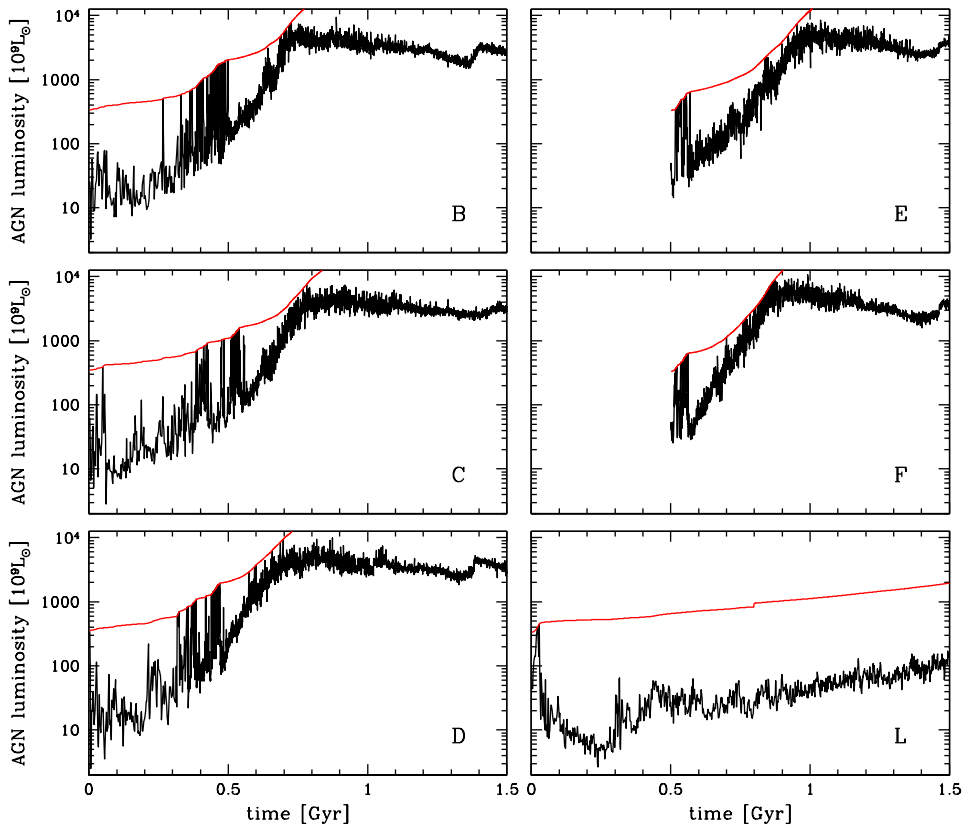}
\caption{
AGN luminosity vs. time for gas-normal galaxies: Run B (thermal feedback),
Run C ($f_{\rm kin}=0.1$),
Run D ($f_{\rm kin}=0.2$), Run E (delayed AGN, $f_{\rm kin}=0.1$),
Run F (delayed AGN, $f_{\rm kin}=0.2$), and Run L (no bar, $f_{\rm kin}=0.2$).
The red lines show the Eddington luminosity.}
\label{feedback}
\end{center}
\end{figure*}

Figure~\ref{feedback} shows the evolution of the AGN luminosity
for gas-normal
galaxies. Remember that only a fraction $\epsilon_r\epsilon_c=0.005$
of the energy released actually couples to the ISM in the form of feedback.
Runs~B, C, and D follow a similar pattern. The luminosity
steadily increases by a factor of order 300, reaching its peak value
$L_{\rm AGN}\sim6\times10^{12}L_\odot$ at $t\sim750\,\rm Myr$. Afterward,
$L_{\rm AGN}$ slowly decreases, even though the black hole mass 
keeps increasing. The removal of the gas in the central region reduces the 
factor $\rho_\infty$ in equation~(\ref{bondi}), and this effect eventually
dominates over the increase in $M_{\rm BH}$. In Runs~E and F, AGN accretion
and feedback is delayed until $t=500\,\rm Myr$, and this in turn delays
the time when the AGN luminosity reaches its peak, around $t=950\,\rm Myr$.
Except for that, the overall evolution of $L_{\rm AGN}$ is similar to the ones
for Runs B, C, and D. We notice large, sudden fluctuations in $L_{\rm AGN}$,
by factors of 20 or more. These fluctuations take place between 
$t=200\,\rm Myr$ and $t=500\,\rm Myr$ in Runs~B, C, and D, and soon after
AGN turn-on in Runs~E and F. This corresponds to a period of rapid black hole
growth. During that period, the gas accreting onto the black hole can
be quite clumpy, leading to sudden variations in $\rho_\infty$ in 
equation~(\ref{bondi}). When a dense clump of gas falls onto the black hole,
$\dot M_{\rm BH}$ can increase by a large factor, and the AGN luminosity is
then given by the Eddington limit (see eq.~\ref{LAGN}). This effect becomes
less important at later times because the gas is less dense, and a more
diffuse gas is less likely to fragment into dense clumps. Notice, however,
the peak at $t=630\,\rm Myr$ in Run~B.

With so little gas in the central region, AGN feedback in unbarred galaxies is
greatly reduced. The bottom right
panel of Figure~\ref{feedback} shows the AGN
luminosity for Run~L. After an initial peak caused by the accretion of gas
that was located near the black hole in the initial conditions, the luminosity
increases slowly. Unlike in barred galaxies, the luminosity never reaches
a peak because the accretion rate is too low to exhaust the supply of gas.
Throughout the evolution, the luminosity remains consistently an order
of magnitude below the Eddington limit, and by $t=1.5\,\rm Gyr$, it is a
factor of 30 lower than in Run~D.

\begin{table}
  \centering
  \begin{minipage}{140mm}
    \caption{Central gas mass and black hole mass}
    \begin{tabular}{@{}cccccc@{}}
      \hline
      Run & $t_{\rm peak}[{\rm Gyr}]{}^a$ & $M_{\rm gas}^{\rm peak}\,{}^b$ &
      $M_{\rm gas}^{\rm final}\,{}^c$ & $|\Delta M_{\rm gas}|\,{}^d$
      & $M_{\rm BH}{}^e$ \\
      \hline
      B & 0.44 & 1.865 & 0.400 & 1.465 & 0.196 \\
      C & 0.49 & 1.464 & 0.368 & 1.096 & 0.189 \\
      D & 0.45 & 1.752 & 0.361 & 1.391 & 0.189 \\
      E & 0.45 & 1.638 & 0.397 & 1.241 & 0.234 \\
      F & 0.52 & 1.652 & 0.368 & 1.284 & 0.191 \\
      \hline
    \end{tabular}
  \end{minipage}
  \medskip
  \begin{minipage}{84mm}
    $^a$ Time when the central gas mass reaches its peak value.\\
    $^b$ Central gas mass at $t=t_{\rm peak}$.\\
    $^c$ Central gas mass at $t=1.5\,\rm Gyr$.\\
    $^d$ Decrease in central gas mass between $t=t_{\rm peak}$
         and $t=1.5\,\rm Gyr$.\\
    $^e$ Central black hole mass at $t=1.5\,\rm Gyr$.\\
    All masses are in units of $10^9[M_\odot]$.     
  \end{minipage}
\end{table}

In Table~2, we compare the final mass of the black hole to the mass
of central gas removed during the period of black hole growth.
$M_{\rm gas}^{\rm peak}$
is the maximum value of the central gas mass, determined from
Figure~\ref{gas_centre}, and $t_{\rm peak}$ is the time when this peak
value is reached. $M_{\rm gas}^{\rm final}$ and $M_{\rm BH}$ are the
masses of the central gas and black hole, respectively, at $t=1.5\,\rm Gyr$.
$|\Delta M_{\rm gas}^{\rm final}|=|M_{\rm gas}^{\rm final}
-M_{\rm gas}^{\rm peak}|$ is the mass of central gas removed. Note that the
black hole mass at $t=t_{\rm peak}$ is negligible compared to the final mass,
for all runs. In all cases,  $|\Delta M_{\rm gas}|$ exceeds $M_{\rm BH}$
by a factor of five or more, implying that most of the gas removed from the
central region is not accreted by the black hole, but instead converted to
stars.  We note that in Figure~\ref{gas_centre}, the evolution of the
central gas mass is essentially
the same for Run A, which has no AGN, as in the other runs.

\subsection{Star Formation History}

\begin{figure}
\begin{center}
\includegraphics[scale=0.50]{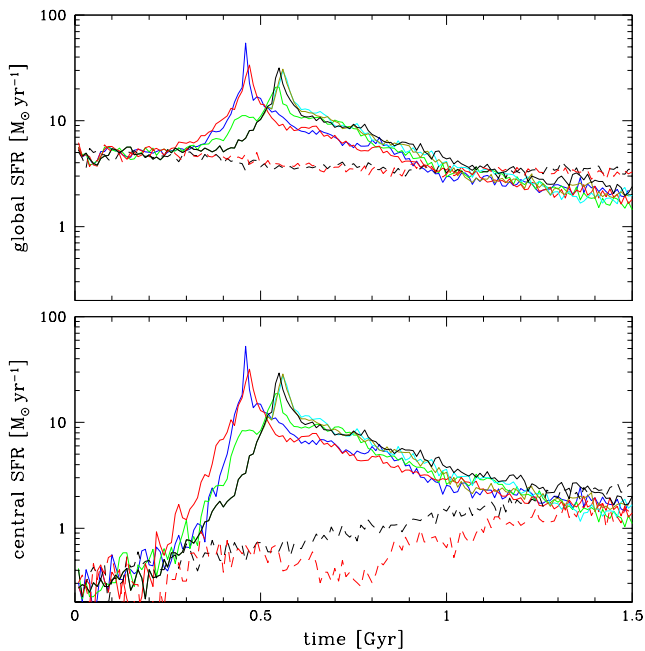}
\caption{
Star formation rate vs. time for gas-normal galaxies. Top panel: entire galaxy;
bottom panel: central $1\,\rm kpc$ region.
Solid black lines: Run A (No AGN);
blue lines: Run B (thermal 
feedback); green lines; Run C ($f_{\rm kin}=0.1$); solid red lines:
Run D ($f_{\rm kin}=0.2$); cyan lines: Run E (delayed AGN, $f_{\rm kin}=0.1$);
gold lines: Run F (delayed AGN, $f_{\rm kin}=0.2$);
dashed black lines: Run K (no bar, no AGN);
dashed red lines: Run L (no bar, $f_{\rm kin}=0.2$).
Note that gold and cyan lines
are identical to black line at $t<0.5\,\rm Gyr$.}
\label{SFR_AF}
\end{center}
\end{figure}

We first consider the build-up of the stellar mass in gas-normal galaxies.
Figure~\ref{SFR_AF} 
shows the SFR vs. time for the entire galaxy (top panel) and inside the
central $1\,\rm kpc$ region (bottom panel).
In the absence of an AGN (Run A, black line), the global
SFR is essentially constant
until $t=400\,\rm Myr$, after the bar has formed. Then, star formation
rapidly increases, and reaches a peak at $t=580\,\rm Myr$. Afterward,
the SFR slowly decreases as the supply of gas available for forming 
new stars gets depleted.
This is fully consistent with the results presented in Papers~I and II.
The presence of an AGN (Runs~B, C, and~D) has no significant effect
on the SFR until $t=300\,\rm Myr$. The central
SFR is small at $t<300\,\rm Myr$, hence most of the stars form in regions
that are too far from the centre to be affected by AGN feedback. Also,
the accretion rate is still relatively small at this time, as
Fig.~\ref{feedback} shows.
After $t=300\,\rm Myr$, star formation in the central region becomes
important, and the SFR increases faster with simulations with feedback.
Looking at Figure~\ref{bar}, we see that $t=300\,\rm Myr$ is roughly
the time when the bar strength in Runs B and C starts increasing faster
than in Run A (the strength of the bar in Run D is already larger
at that time).
After the peak is reached, the SFR decreases
roughly at the same rate as for Run A, and stays about 30\% lower
than Run A from $t=800\,\rm Myr$ until the end of the simulation.
Note that at these late stages, star formation occurs almost
exclusively in the central region, which explains the similarity
of the curves in the top and bottom panels at late times.
In all cases, the peak SFR is reached $\sim70\,\rm Myr$ after
the peak bar strength is reached. This suggests that the differences
in SFR at radii of $1\,\rm kpc$ and larger
are mostly a result of the different bar strengths, and that
any direct effect of AGN feedback must take place at smaller radii.

Delaying the turn-on of AGN feedback greatly affects the star
formation history.
In Runs E and F, the AGN is turned on at $t=500\,\rm Myr$, at a time
when the bar is formed and
star formation in the central regions is well under way. AGN feedback
has essentially no effect, because the black hole mass is still at 
$M_{\rm BH}=10^6\msun$, resulting in a weak AGN luminosity. It takes another
$500\,\rm Myr$ before the AGN luminosity in Runs~E and F catches up with 
Runs~C and D, and by that time star formation is well passed its peak.
As a result, the time and height of the SFR peaks are
essentially the same for Runs A, E, and F, and the late-time evolutions of
the SFR are also the same. 

The dashed lines in Figure~\ref{SFR_AF}
show the SFR for unbarred galaxies. The global SFR
slowly decreases with time, dropping by a factor of
about 2 after $1.5\,\rm Gyr$, and is
not affected by feedback. The central SFR increases slowly and more 
irregularly, and the effect of feedback is small except for a brief period
around  $t=800\,\rm Myr$.

\begin{figure*}
\begin{center}
\includegraphics[scale=0.8]{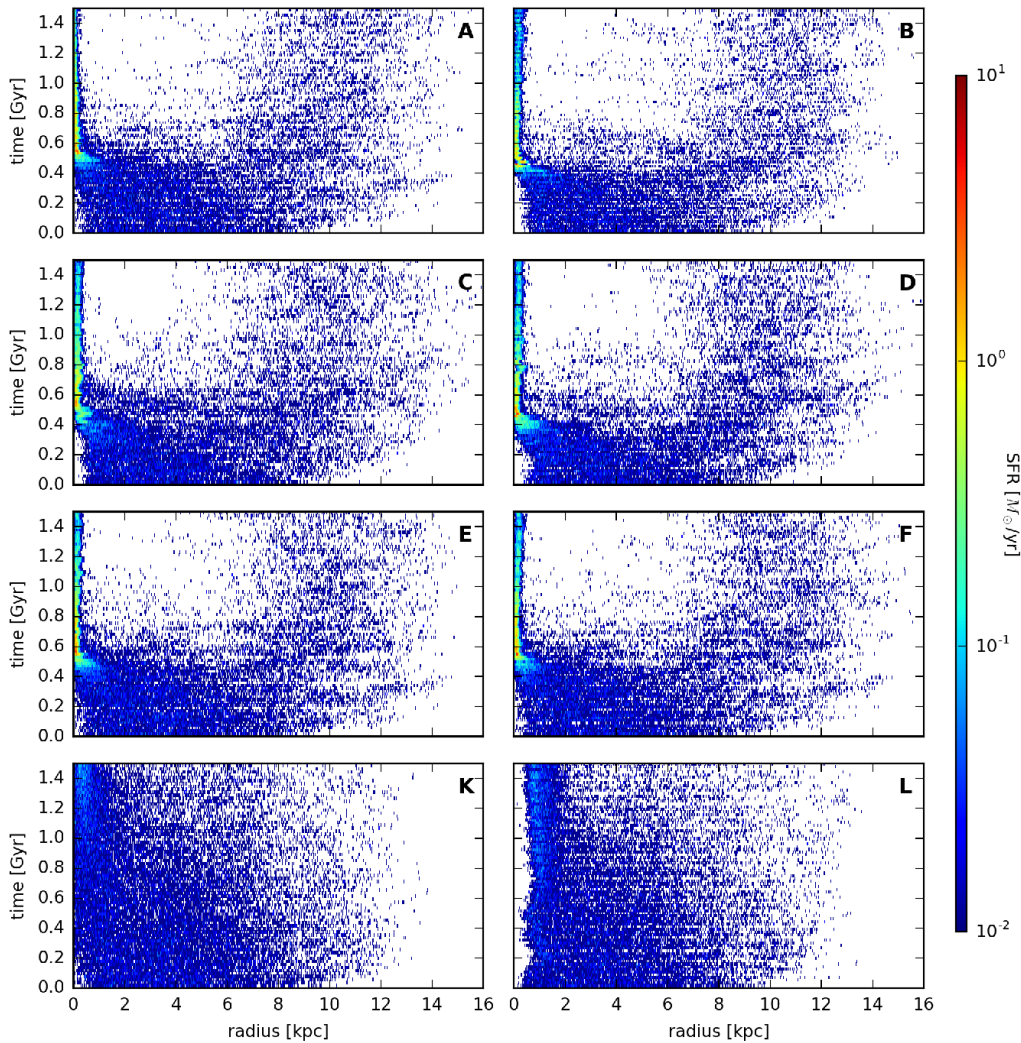}
\caption{Star formation maps for
for gas-normal galaxies, showing entire galaxy:
Run A (No AGN), Run B (thermal feedback), Run C ($f_{\rm kin}=0.1$),
Run D ($f_{\rm kin}=0.2$), Run E (delayed AGN, $f_{\rm kin}=0.1$),
Run F (delayed AGN, $f_{\rm kin}=0.2$),
Run K (no bar, no AGN), and Run L (no bar, $f_{\rm kin}=0.2$).}
\label{Map_SF1}
\end{center}
\end{figure*}

\begin{figure*}
\begin{center}
\includegraphics[scale=0.8]{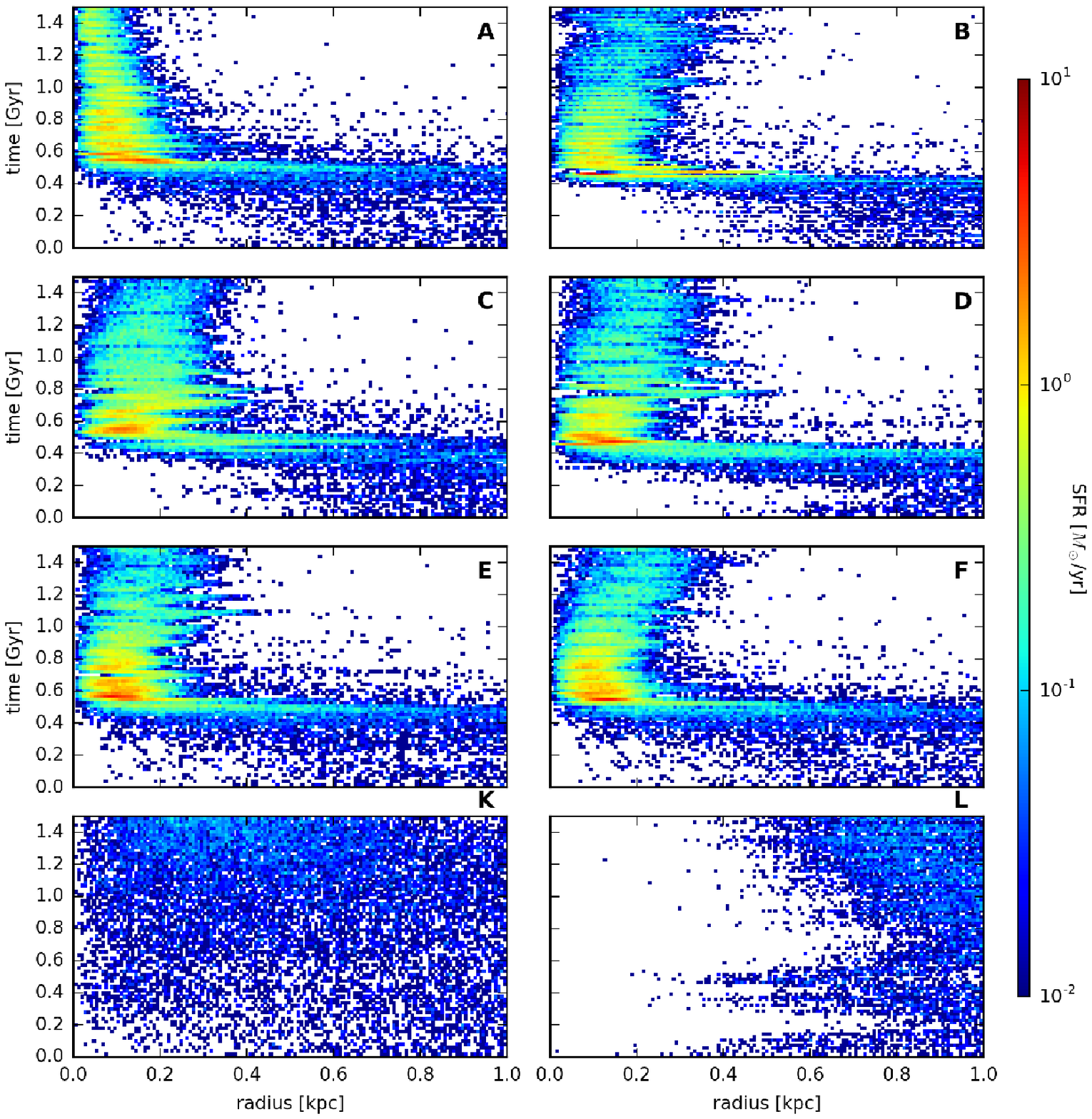}
\caption{Star formation maps for
for gas-normal galaxies, showing central $1\,\rm kpc$ region:
Run A (No AGN), Run B (thermal feedback), Run C ($f_{\rm kin}=0.1$),
Run D ($f_{\rm kin}=0.2$), Run E (delayed AGN, $f_{\rm kin}=0.1$),
Run F (delayed AGN, $f_{\rm kin}=0.2$),
Run K (no bar, no AGN), and Run L (no bar, $f_{\rm kin}=0.2$).}
\label{Map_SF2}
\end{center}
\end{figure*}

Figure~\ref{Map_SF1} 
shows ``spacetime maps'' of the star formation history. Each pixel 
shows the average SFR in solar masses per year for
a particular radial bin over a particular time interval.
Looking at Run~A, stars initially form at all radii below 
$9\,\rm kpc$ except at the very centre where the amount of gas
available is initially low.
As time goes on, angular momentum is redistributed inside the disc. After
$t=500\,\rm Myr$, we clearly have two separate regions of active
star formation: an inner region ($r<2\,\rm kpc$) where stars are
forming inside the bar, and an outer region ($r>6\,\rm kpc$)
where stars are forming in the disc, mostly inside the spiral arms.
Figure~\ref{Map_SF2} shows a zoom-in of the central $1\,\rm kpc$ region.
During the early stages prior to the formation of the bar
($t<0.4\,\rm Gyr$), there is a significant suppression of star formation in
the centre at early time, because the gas density is too low to trigger
star formation, and gas gets
instead accreted by the black hole. In Runs~B, C, and D, and also in Run~L,
the gas that does
not get accreted is then pushed to larger radii by the black hole feedback.
In Runs~A, E, and F, there is no feedback at early time, allowing a
limited amount of star formation down to small radii, and the effect is much
larger in Run~K because the central gas density is initially larger.

At $t>0.4\,\rm Gyr$, the bar has formed in all simulations A through F,
and the situation changes drastically.
For Run~A, we clearly see
the peak in star formation at $580\,\rm Myr$ (red area).
As time goes on, star formation gets more centrally concentrated. 
This is consistent with the results of Paper~I. The ``classic'' scenario
for the evolution of barred galaxies states that the bar drives gas
toward the central region, where it is then converted into stars.
We showed in Paper~I that stars actually form along the entire length
of the bar, and star formation gets more concentrated because the
gaseous component of the bar contracts with time.

The effect of AGN feedback on central star formation is clearly visible.
Comparing Runs~B, C, and~D with Run~A, we see that
in the presence of AGN feedback, the region
of active star formation no longer contracts with time, and in Run~B it even
expands. At $t=1.5\,\rm Gyr$, central star formation is concentrated
in the inner $100\,\rm pc$ in Run~A, but reaches $400\,\rm pc$ in
Runs~B, C, and~D. The algorithm turns gas particles into star particles 
when the hydrogen gas density reaches
a threshold value $n_{\rm th}^{\phantom1}=1\,\rm cm^{-3}$. In the absence
of AGN feedback, the gas inside the bar reaches that threshold density
``on its own,'' as the bar contracts with time. In the presence of feedback,
gas inflowing along the bar toward the centre collides head-on with gas
being pushed outward by the AGN, causing a rapid increase in density.
This is consistent with the model of \citet{if12}, where an AGN can trigger
star formation by driving a shell of material outward. In our simulations,
this results in star formation being pushed to larger radii, starting earlier,
and being more violent.

In Runs~E and~F, feedback is delayed until $t=0.5\,\rm Gyr$. At that time,
star formation is well under way. The addition of feedback does not
significantly affect the star formation rate, since the gas is already
very dense. However, feedback does push the central star formation to larger
radii, as it does in Runs~B, C, and~D.

In Run~K, where no bar is present to transfer angular momentum,
star formation takes place at all radii up to $r=12\,\rm kpc$. Comparing
Runs~K and L, we see that feedback pushes central star formation out to 
larger radii, as it does for barred galaxies. In Run~L, star formation is
nearly suppressed at radii $r<0.4\,\rm kpc$, and greatly reduced
at radii $0.4<r<0.8\,\rm kpc$.
Comparing Runs~D and L, we see that the 
ability to push star formation outward is much higher in the absence of a bar.
In Run~L, outflowing gas from the AGN is pushing on gas that is nearly 
standing still, while in Run~D it is colliding with gas that is 
moving inward. Not only it is more difficult to push inflowing gas, 
but the resulting compression triggers star formation earlier.

In all cases, we find that the effects of AGN feedback are confined
to a region much smaller than the ``centre,'' which we define as the
central $1\,\rm kpc$. This explains why the differences in SFR
seen in Figure~\ref{SFR_AF} at early times
are mostly due to differences in bar strength.

\begin{figure}
\begin{center}
\includegraphics[scale=0.45]{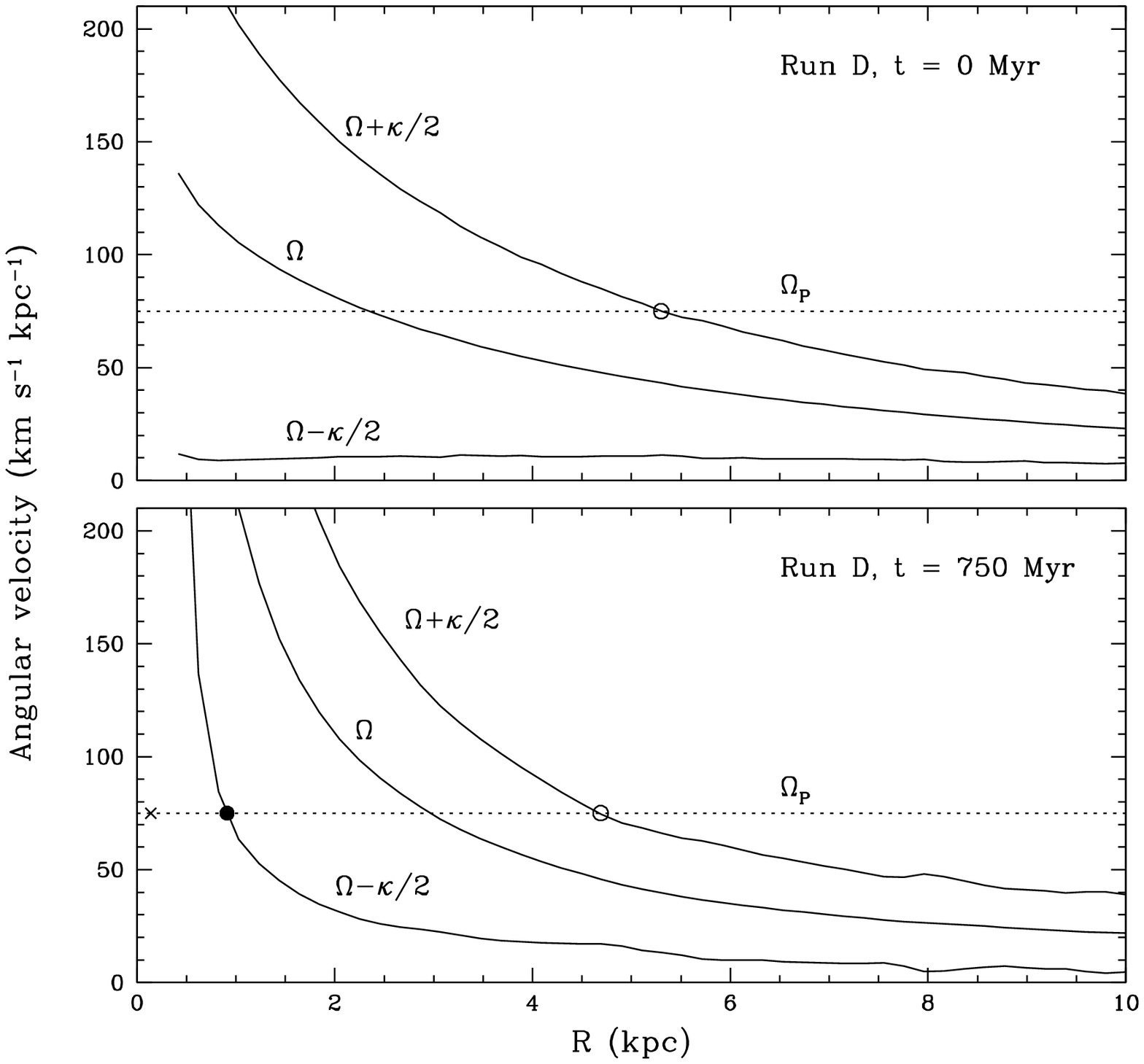}
\caption{Resonances in the galactic disc, for Run D.
Top panel: initial time; bottom panel: at $t=750\,\rm Myr$.
Solid lines: $\Omega+\kappa/2$, $\Omega$, and
$\Omega-\kappa/2$ versus radius $R$, as indicated, where $\Omega$ is the
circular frequency and $\kappa$ is the epicycle frequency.
Dotted lines: pattern rotation velocity $\Omega_{\rm p}$.
The cross, solid circle, and open circles show the location of
the star formation ring (inferred from Fig.~\ref{Map_SF2}), the inner
Lindblad resonance, and the outer Lindblad resonances, respectively.}
\label{resonances}
\end{center}
\end{figure}

To further demonstrate that the formation of a star formation ``ring'' is
a result of AGN feedback, and not a dynamical effect, we calculated the
rotation velocity profiles, to search for resonances.
Figure~\ref{resonances} shows the rotation velocity profiles for Run~D,
at the initial time and at time $t=750\,\rm Myr$, after the bar is well-formed.
$\Omega$, $\kappa$, and $\Omega_{\rm p}$ are the circular frequency, epicycle
frequency, and pattern rotation velocity,
respectively. At $t=0$, we find an outer
Lindblad resonance at $R=5.2\,\rm kpc$, but no inner Lindblad resonance.
This allows the gas to flow toward the centre unimpeded, and accumulate
in the central region. At later time, an inner Lindblad resonance appears
at radius $R=0.91\,\rm kpc$ (solid circle in bottom panel of
Fig.~\ref{resonances}). Such resonance could lead to the accumulation of
gas at that radius, but this does not coincide with the location of the
star formation ring. From Figure~\ref{Map_SF2}, we estimate that the
star formation ring, at $t=750\,\rm Myr$, for Run~D, is centred at about
$R=0.14\,\rm kpc$ (cross in bottom panel of
Fig.~\ref{resonances}). Therefore, there is no evidence that resonance play
any significant role in the formation of a star formation ring.

\subsection{Gas Fraction}

\begin{figure}
\begin{center}
\includegraphics[scale=0.50]{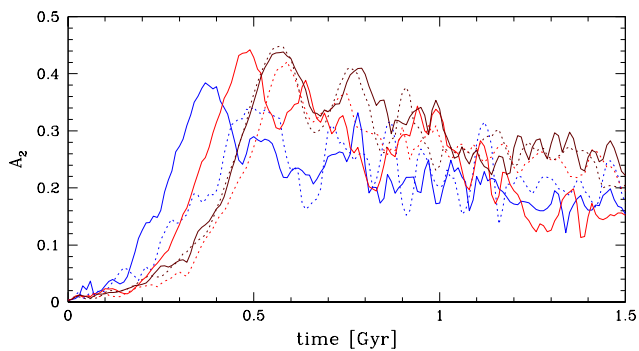}
\caption{
Bar strength vs. time for galaxies with no AGN (dotted lines)
and with AGN feedback ($f_{\rm kin}=0.2$, solid lines). 
Blue lines: Runs A and D (gas-normal galaxies); red lines: Runs 
G and H (gas-poor galaxies); brown lines: Runs I and J
(gas-very-poor galaxies).}
\label{bar2}
\end{center}
\end{figure}

We now focus on the effect of varying the initial gas fraction
in barred galaxies.
In Runs~A through~F (the ``gas-normal galaxies''), the initial gas fraction
$f_{\rm gas}=0.192$, in accordance with equation~(\ref{mgasinit}).
Runs~G and~I, and Runs~H and~J, are similar to Runs~A and~D, respectively
(same total mass, and for Runs~H and~J, same AGN feedback prescription
as Run~D), but in Runs~G and~H (the ``gas-poor'' galaxies), the initial gas
fraction $f_{\rm gas}=0.096$, and in Runs~I and~J 
(the ``gas-very-poor'' galaxies), $f_{\rm gas}=0.048$.

The evolution of the bar strength is shown in Figure~\ref{bar2}.
In this figure and the following ones, we use dotted lines
for the runs with no AGN (A, G, I) and solid lines for the runs
with AGN (D, H, J). Blue, red, and brown lines show gas-normal,
gas-poor, and gas-very-poor
galaxies, respectively. When AGN feedback is present there is perhaps a
trend of bar formation becoming more delayed with decreasing gas fraction,
but this trend is not clear in the runs without AGN feedback, and may not
be significant. Once the bar has reached its maximum strength, the
subsequent evolution is qualitatively the same for all runs: a strong
oscillation superposed on a slow decay. The final values of $A_2$ are
all in the range 0.15--0.25, with no obvious correlation with gas fraction.
We notice again a non-monotonicity, with the bar forming slightly earlier
for Run I (dotted brown line) than Run G (dotted red line).

\begin{figure}
\begin{center}
\includegraphics[scale=0.50]{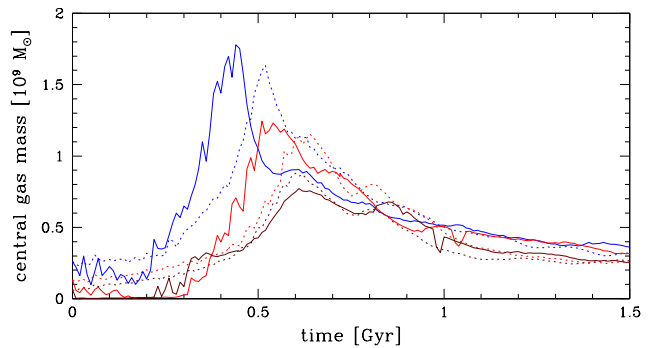}
\caption{
Gas mass in the central region vs. time for galaxies with no AGN (dotted lines)
and with AGN feedback ($f_{\rm kin}=0.2$, solid lines).
Blue lines: Runs A and D (gas-normal galaxies); red lines: Runs 
G and H (gas-poor galaxies); brown lines: Runs I and J (gas-very-poor galaxies).
}
\label{gas_centre2}
\end{center}
\end{figure}

Figure~\ref{gas_centre2} shows the evolution of the gas mass in the central
$1\,\rm kpc$, for Runs A, D, and G--J. As the gas fraction is reduced, the
central gas mass grows slower, peaks at a lower value, and reaches that
peak value later. Gas is a dissipational component that can strengthen
instabilities \citep{elmegreen11},
and larger gas fractions can help drive gas inwards.
Once the peak is passed, the late-time evolution is similar.
The amount of gas left in the centre is slightly reduced when
the initial gas fraction is reduced by a factor of 2.

\begin{figure}
\begin{center}
\includegraphics[scale=0.50]{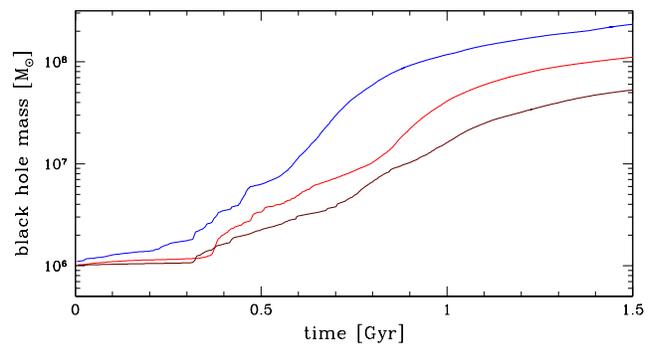}
\caption{
Black hole mass vs. time for galaxies with AGN feedback 
($f_{\rm kin}=0.2$). Blue line: Run D (gas-normal galaxy); red line: Run 
H (gas-poor galaxy); brown line: Run J (gas-very-poor galaxy).
}
\label{mass_BH2}
\end{center}
\end{figure}

Figure~\ref{mass_BH2} shows the evolution of the black hole mass
for Runs~D, H, and~J.  In Run~D, the black hole mass starts growing
immediately, albeit at a slow rate until $t=300\,\rm Myr$. In Runs~H
and~J, the black hole mass is nearly constant until $t=300\,\rm Myr$.
The bar is initially stronger in Run~D, driving more gas to the centre.
At late time, the evolution is similar for all runs, and the final
values at $t=1.5\,\rm Gyr$ are proportional to the initial values
of $f_{\rm gas}$, differing by a factor of 2 between Runs~D and~H,
and a factor of 2 between Runs~H and~J.

\begin{figure}
\begin{center}
\includegraphics[scale=0.80]{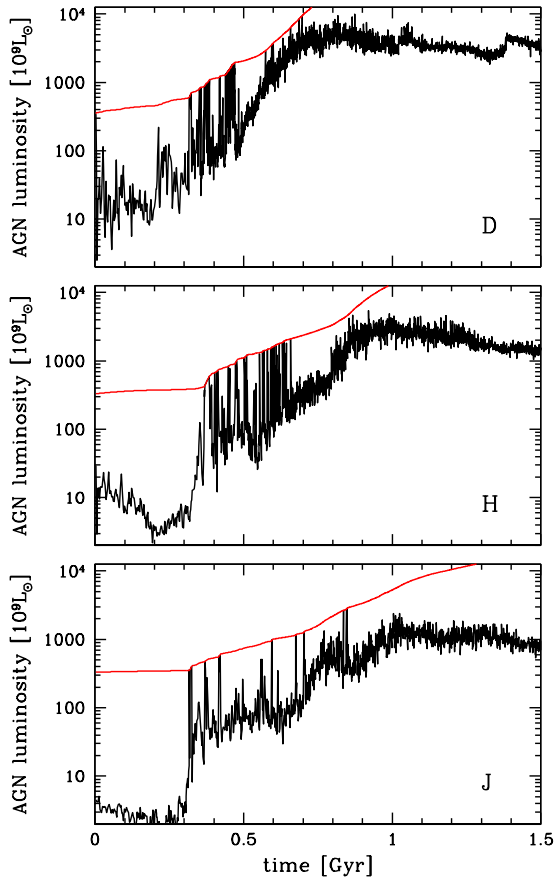}
\caption{
AGN luminosity vs. time for barred galaxies with 
$t_{\rm AGN}=0\,\rm Gyr$
and $f_{\rm kin}=0.2$: Run D ($f_{\rm gas}=0.192$),
Run~H ($f_{\rm gas}=0.096$), and Run~J ($f_{\rm gas}=0.048$), as indicated.
The red lines show the Eddington luminosity.
}
\label{feedback2}
\end{center}
\end{figure}

Figure~\ref{feedback2} shows the AGN luminosity, vs. 
time, for Runs~D, H, and J. As Figure~\ref{mass_BH2} showed, reducing
the gas fraction reduces the black hole mass $M_{\rm BH}$. Since the
Bondi accretion rate and the Eddington accretion rate both 
increase with $M_{\rm BH}$,
the AGN luminosity decreases with decreasing gas fraction. The peak luminosity
is reached later for lower gas fractions, and once the peak is passed, the
luminosity decreases slowly, for all runs, as the factor $\rho_\infty$ in
equation~(\ref{bondi}) drops.

\begin{figure}
\begin{center}
\includegraphics[scale=0.50]{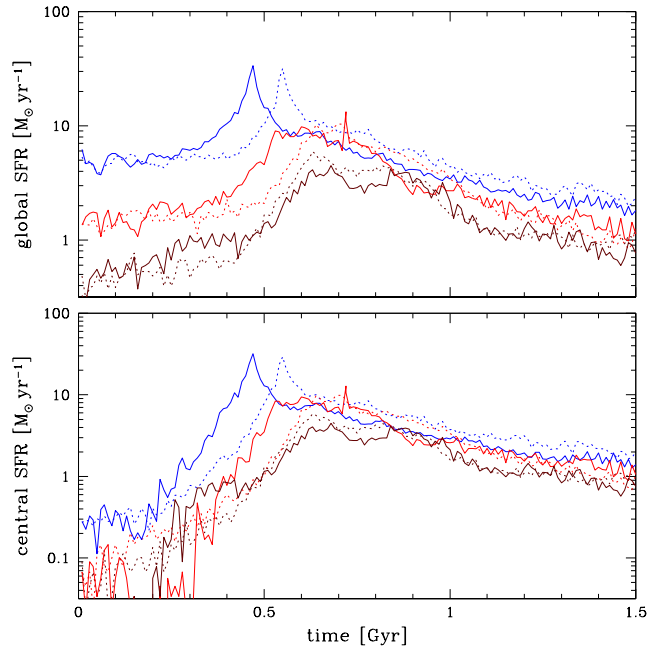}
\caption{Star formation rate vs. time for galaxies with no AGN (dotted lines)
and with AGN feedback ($f_{\rm kin}=0.2$, solid lines).
Top panel: entire galaxy;
bottom panel: central $1\,\rm kpc$ region.
Blue lines: 
Runs A and D (gas-normal galaxies); red lines: Runs 
G and H (gas-poor galaxies); brown lines: Runs I and J (gas-very-poor galaxies).
}
\label{sfr2}
\end{center}
\end{figure}

Figure~\ref{sfr2} shows the evolution of the SFR,
for Runs A, D, and G--J.
The SFR drops significantly with gas fraction, both for galaxies with an AGN
(solid lines) and without (dotted lines). Not only the gas
density is reduced in gas-poor galaxies, directly affecting star formation 
efficiently, but this reduction in density also increases the cooling time 
of the gas, since the cooling rate scales like the square of the density.
This is reflected both in the global and the central SFR. With a lower
gas fraction, the SFR rises slower, peaks at a lower value, and reaches
this peak later. Comparing the bottom panel of Figure~\ref{sfr2}
with the top panel of Figure~\ref{gas_centre2}, we see that the central
SFR closely follows the central gas mass. 

\begin{figure*}
\begin{center}
\includegraphics[scale=0.72]{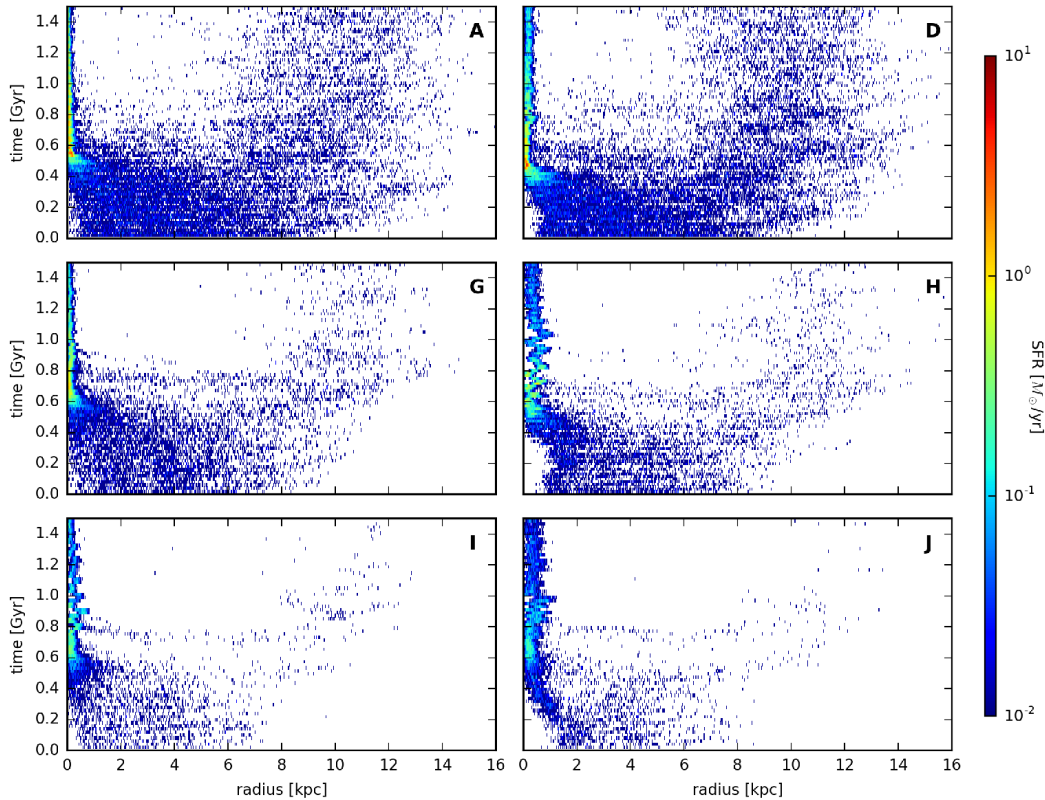}
\caption{Star formation maps for gas-normal galaxies (top row),
gas-poor galaxies (middle row), and gas-very-poor galaxies (bottom row),
showing entire galaxy.
Left column: with no AGN feedback.
Right column: with AGN feedback ($f_{\rm kin}=0.2$).}
\label{Map_SF3}
\end{center}
\end{figure*}

\begin{figure*}
\begin{center}
\includegraphics[scale=0.72]{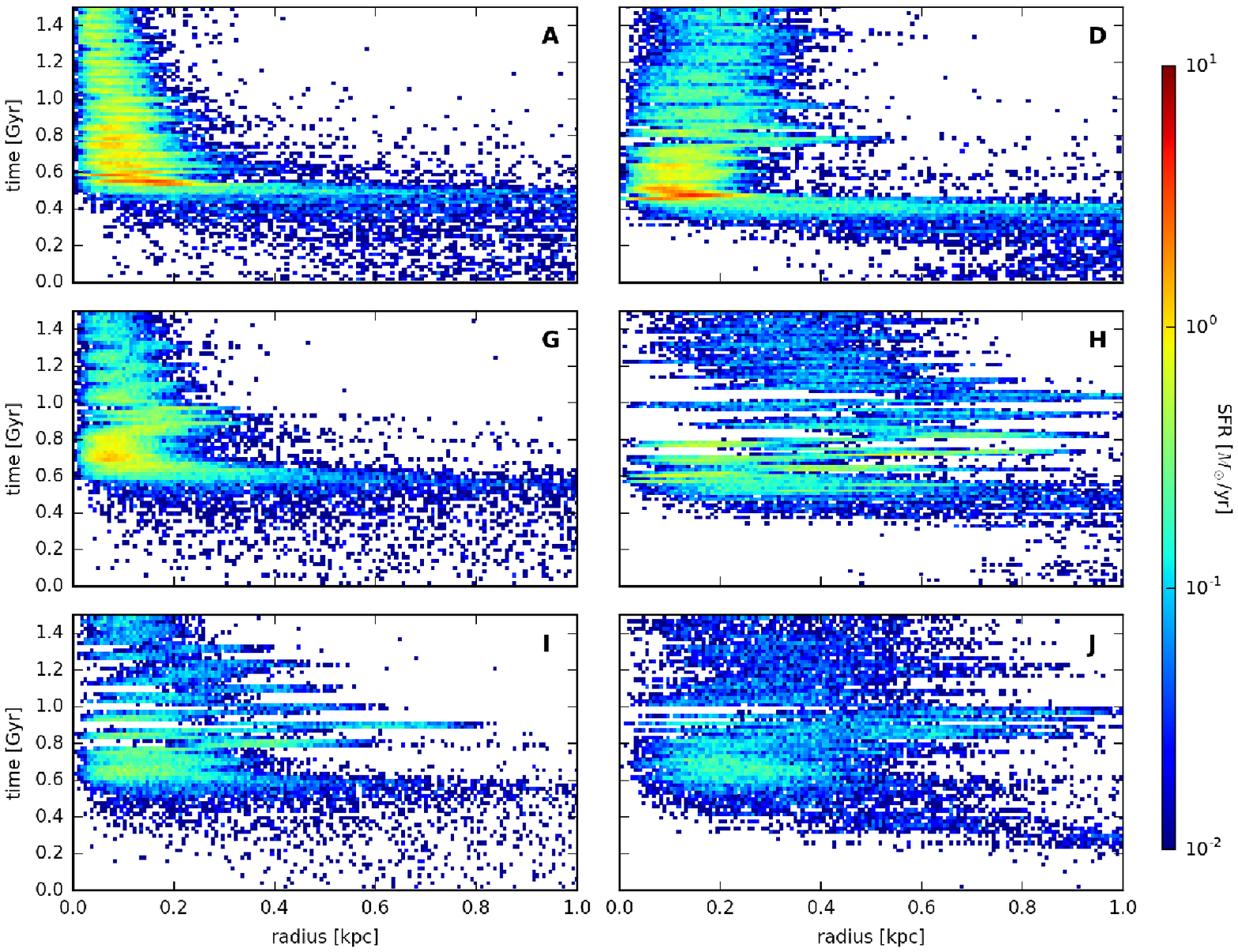}
\caption{Star formation maps for gas-normal galaxies (top row),
gas-poor galaxies (middle row), and gas-very-poor galaxies (bottom row),
showing central $1\,\rm kpc$ region.
Left column: with no AGN feedback.
Right column: with AGN feedback ($f_{\rm kin}=0.2$).}
\label{Map_SF4}
\end{center}
\end{figure*}

Figures~\ref{Map_SF3} and \ref{Map_SF4} 
show spacetime maps of the star formation history,
for the entire galaxy and inside the
central region, respectively, for Runs~A, D, and G--J. Reducing the initial
gas mass reduces star formation activity, particularly in the outer regions.
Indeed, star formation in the outer regions is nearly absent in 
gas-very-poor galaxies. AGN feedback pushes star formation to larger radii
(right column of Fig.~\ref{Map_SF4}). Lowering the gas fraction
leads to competing effects: on one hand the AGN luminosity is lower, but on
the other hand the more diffuse gas is more susceptible to AGN feedback.
Going from Run~D (gas-normal) to Run H (gas-poor),
the nuclear star formation ``cavity'' gets bigger:
star formation is pushed outward, indicating that
the second effect dominates. Going from Run~H (gas-poor) to
Run~J (gas-very-poor), however,
does not affect the location of central star formation very much, as the two
effects mostly cancel each other.

\section{DISCUSSION}

We considered isolated disc galaxies in which the bar forms by instability.
Other processes, such as mergers and tidal interactions, can also lead
to the formation (or destruction)
of a bar. In this study, we are focussing on the
interplay between the bar and the AGN activity, and their impact
on star formation, and not on the bar formation process itself.
We see bar instability as a tool that enables us
to numerically generate galaxies with bars. The alternative
would have been to generate initial conditions in which the bar
was already present, but this approach might produce initial
conditions that are somewhat artificially unstable. By allowing the bar
to form by instability, we ensure that the system is relaxed by
the time the bar forms.

We discovered, somewhat surprisingly, that the bar strength $A_2$
reached its peak value earlier in simulations with AGN feedback.
It appears that feedback speeds up the growth of bar instability,
a phenomenon also noticed by \citet{spinosoetal17}.
The origin of this trend is
certainly worth investigating further, but this is not
the goal of the present study. In all simulations with bars, the
peak values of $A_2$ are similar, and the post-peak evolution of
$A_2$ is also similar. The differences between simulations
do not result from different bar strengths. As long as we regard bar
instability as a mere tool to generate initial conditions, the pre-peak
evolution of the bar is not relevant to this study, and we will
postpone a detailed investigation
of the effect of AGN feedback on bar instability to further work.

Since our goal was to assess the relative importance of
positive and negative feedback, we were not interested in situations where
AGN feedback blows all the gas outside of the galaxy. For this reason, we
chose a regime of
galaxy mass large enough for AGN feedback to dominate over
SNe feedback, yet low enough to retain most of its gas. Also, our algorithm
does not allow for the presence of anisotropic feedback (that is, jets),
since the region near the black hole that would be responsible for
making the ejecta anisotropic is not resolved by the algorithm.
Still, our numerical models provide numerical experiments in the regime
of relatively weak AGN feedback which only affects the central region of
the galaxy. Our comparison of models with different AGN strengths,
barred vs. unbarred galaxies, and different gas fractions provides
qualitative trends of the effect of AGN feedback in the case of weak
AGN feedback systems.

In all simulations with feedback, we use an AGN feedback model
based on \citet{wt13a}.
As noted by 
\citet{baraietal14}, the Bondi accretion rate assumes a steady, adiabatic,
spherically symmetric inflow, which is unlikely to be the case for a
supermassive black hole, and that a more realistic accretion rate may differ
from the Bondi accretion rate by orders of magnitude. Furthermore, the
density at the Bondi radius is only resolved at high resolution, and so it
is necessary to incorporate additional factors to correct for this.
Additionally, the angular momentum of the inflowing gas has an impact on the
accretion rate \citep{pnk11,wt13b}, but this is difficult to capture at
galaxy-scale and cosmological resolutions.
Considering these uncertainties,
we decided to base our algorithm on the WT model of \citet{wt13a},
because it is simple and numerically stable. Finally, we note that
our AGN model produces luminosities of order $10^{12}L_\odot$ even
in the later phase. This likely violates the number fraction of the luminous
AGN in this type of galaxies. Therefore, our model overestimates the
accretion rate in the late phase, and highlights that this is a challenge
for current AGN models.

As in Papers I and II, we have only considered isolated galaxies,
ignoring accretion from the intergalactic medium,
and merger with other galaxies.
These processes could affect the dynamics of the
bar, and also affect the post-starburst evolution of barred galaxies, by
replenishing the supply of gas depleted by star formation
(see, however, \citealt{ellisonetal15a}).
Our simulations are most relevant to galaxies located at
sufficiently low-redshift
that most of the mass assembly is completed
(e.g. \citealt{lcs12}).
We will consider the effects of accretion and mergers in future work.

\section{SUMMARY AND CONCLUSION}

The goal of this study was to determine if AGN feedback has a positive
or a negative effect on star formation in galaxies,
in a regime of relatively weak AGN feedback which only affects
the central regions of the galaxies. 
We have performed a series of 12 simulations of equal-mass barred 
and unbarred disc
galaxies, with various initial gas fractions and various prescriptions
for AGN feedback. We focussed on barred galaxies because of the ability
of a bar to channel gas toward the centre of galaxies and feed
an AGN. Our main results are the following:

\begin{itemize}

\item In all simulations, the bar forms and grows by instability.
The presence of an AGN can hasten the onset
of bar growth, and the bars reach their peak strength earlier.
But eventually all bars settle to a similar strength,
independent of AGN feedback prescription or initial gas fraction.
  
\item AGN feedback initially has a strong
positive effect on star formation. The SFR
increases faster, and peaks earlier.
In the central regions
(smaller than $1\,\rm kpc$), this difference can
be attributed directly to AGN feedback. At larger radii, the difference
results from delays in the formation of the bar.
After most of the available gas has
been converted into stars, the evolution of the SFR is essentially the
same for all runs.
Eventually, the SFR in simulations without feedback catches up, and at
late time the total stellar mass is slightly larger than
in simulations with feedback.

\item The effect of AGN feedback is most important when it can act
before star formation becomes significant. If the turn-on of AGN feedback
is delayed until after star formation in the central region is well under 
way, the effect on the SFR is small.

\item Feedback greatly affects the dynamics of the central region.
In the absence of feedback, star formation peaks at the very centre of the
galaxy, where the gas density is highest. Feedback
pushes gas outward, where it supersonically collides with gas
inflowing along the bar. As a result, star formation starts earlier, is
more dramatic, and is pushed to larger radii.

\item When the initial gas fraction is reduced, the formation of the bar 
is delayed. This in turns delays star formation and reduces the peak
value of the SFR. The final values of the global and central
stellar masses $M_*$ decrease with decreasing initial gas fraction, 
and are hardly affected by AGN feedback.

\item The bar plays a significant role in the evolution of the galaxies,
by driving gas toward the centre where if can form stars and feed the AGN.
By the end of the simulations, the black hole mass was lower by a factor of
30 and the central stellar
mass was lower by a factor of more than 3 in unbarred galaxies compared
to barred ones. Furthermore, the effect of AGN feedback in unbarred galaxies 
is negligible, except in the central $1\,\rm kpc$ region where
star formation is suppressed. In particular, the global
SFR is nearly identical in unbarred galaxies with and without
feedback. Compared with barred galaxies, unbarred galaxies have a much lower
AGN luminosity and the central region contains less gas that is
susceptible to be affected by feedback.

Our final conclusion is that both positive and negative AGN feedback
are present. Feedback suppresses star formation near the central black
hole (negative feedback). The gas is pushed outward where it collides
with inflowing gas, forming a dense ring in which star formation is
enhanced (positive feedback).
In barred galaxies, these two effects mostly cancel out. Stars form
at different locations and times, but in a similar amount.
In unbarred galaxies, negative feedback is more efficient and positive
feedback is less efficient, leading to a net reduction of star formation.
Our results are fully consistent with the
analytical model proposed by \citet{if12}.

\end{itemize}

\section*{acknowledgments}

This research is supported by the Canada Research Chair program and NSERC.
DJW is supported by European Research Commission grant ERC-StG-6771177
DUST-IN-THE-WIND.

%

\label{lastpage}

\end{document}